\documentclass[useAMS,twocolumn,usenatbib]{mn2e}
\pdfoutput=1
\usepackage{times,psfig,epsfig}
\usepackage{rotating, graphicx}
\graphicspath{{figures/}}

\usepackage{color}
\usepackage{amssymb, amsmath}
\usepackage{multirow}
\usepackage{subfigure}

\usepackage[titletoc]{appendix}

\usepackage{enumerate}
\newif\ifAMStwofonts
\AMStwofontstrue

\def\mvir{M_{\rm vir}}
\def\rvir{R_{\rm vir}}

\def\rfive{R_{\rm 500}}

\def\mtwo{M_{\rm 200}}
\def\rtwo{R_{\rm 200}}

\def\msun{\rm{\,M_{\odot}}}

\def\msunh{\,h^{-1}\rm{\,M_{\odot}}}

\def\kms{\rm{\,km\,s^{-1}}}
\def\kpc{\rm{\,kpc}}

\def\mpc{\rm{\,Mpc}}
\def\mpch{\,h^{-1}\rm{\,Mpc}}

\def\agn{{\tt AGN}}
\def\csf{{\tt CSF}}

\def\hi{{\tt Hi}}
\def\me{{\tt Me}}
\def\lo{{\tt Lo}}

\newcommand{\be}{\begin{equation}}
\newcommand{\ee}{\end{equation}}

\newcommand{\chandra}{{\it Chandra}}
\newcommand{\asca}{{\it ASCA}}
\newcommand{\sax}{{\it BeppoSAX}}

\newcommand{\xmm}{{\it XMM-Newton}}

\newcommand{\suzaku}{{\it SUZAKU}}
\newcommand{\athena}{{\it Athena}}

\newcommand{\figref}{Fig.~\ref}

\title[Origin of ICM metallicity in cluster outskirts]{\boldmath The origin of ICM enrichment in the outskirts of present-day galaxy clusters from cosmological hydrodynamical simulations}

\author[V.~Biffi et al.]{V.~Biffi$^{1,2}$\thanks{e-mail: biffi@oats.inaf.it},
  S.~Planelles$^{3}$,
S.~Borgani$^{1,2,4}$, E.~Rasia$^{2}$,
G.~Murante$^2$, D.~Fabjan$^{5,2}$,
\newauthor
M.~Gaspari$^6$\thanks{Einstein and Spitzer Fellow.}
\\~\\
\footnotesize
$^1$ Dipartimento di Fisica dell' Universit\`a di Trieste, Sezione di Astronomia, via Tiepolo 11, I-34131 Trieste, Italy\\
$^2$ INAF, Osservatorio Astronomico di Trieste, via Tiepolo 11, I-34131, Trieste, Italy\\
$^3$ Departamento de Astronom{\'i}a y Astrof{\'i}sica, Universidad de Valencia, c/ Dr. Moliner, 50, 46100 - Burjassot (Valencia), Spain\\
$^4$ INFN --- National Institute for Nuclear Physics, Via Valerio 2, I-34127 Trieste, Italy\\
$^5$ Faculty of Mathematics and Physics, University of Ljubljana, Jadranska 19, 1000 Ljubljana, Slovenia \\
$^6$ Department of Astrophysical Sciences, Princeton University, Princeton, NJ 08544, USA; Einstein and Spitzer Fellow
}

\begin{document}
\maketitle
\begin{abstract}
The uniformity of the intra-cluster medium (ICM) enrichment level in
the outskirts of nearby galaxy clusters suggests that chemical
elements were deposited and widely spread into the intergalactic
medium before the cluster formation. This observational evidence is
supported by numerical findings from cosmological hydrodynamical
simulations, as presented in~\protect\cite{Biffi_2017}, including the
effect of thermal feedback from active galactic nuclei.  Here, we
further investigate this picture, by tracing back in time the spatial
origin and metallicity evolution of the gas residing at $z=0$ in the
outskirts of simulated galaxy clusters.  In these regions, we find a
large distribution of iron abundances, including a component of
highly-enriched gas, already present at $z=2$. At $z>1$, the gas in the
present-day outskirts was distributed over tens of virial radii from
the the main cluster and had been already enriched within
high-redshift haloes.  At $z=2$, about $40\%$ of the most Fe-rich gas at
$z=0$ was not residing in any halo more massive than $10^{11}\msunh$ in the
region and yet its average iron abundance was already $0.4$, w.r.t. the
solar value by~\protect\cite{angr1989}.  This confirms that the in
situ enrichment of the ICM in the outskirts of present-day clusters
does not play a significant role, and its uniform metal abundance is
rather the consequence of the accretion of both low-metallicity and
pre-enriched (at $z>2$) gas, from the diffuse component and through
merging substructures.  These findings do not depend on the mass of
the cluster nor on its core properties.
\end{abstract}


\section{Introduction}
\label{sec:intro}

Galaxy clusters, characterized by deep gravitational potential wells,
retain crucial information on the different physical and dynamical
processes  affecting their stellar, gaseous and dark matter components
\citep[see][for reviews]{kravtsov2012, Planelles_2016}. Therefore, a
thorough understanding of the redshift evolution and the distribution
of the different cluster components is essential to deepen our
comprehension of observed cluster properties.
In this sense, X-ray observations of galaxy clusters
\citep[see][for a review]{Bohringer_2010} have been particularly successful
in providing information on the distributions of density,
temperature and different metal species within the hot intra-cluster medium (ICM).

The distribution of metals mainly results from the formation of
stars in cluster galaxies and from supernova (SN) events.
Later on, once metals have been released, thanks to the
combined action of a number of processes
\citep[such as AGN feedback, galactic winds or ram-pressure
stripping, e.g.][]{Churazov_2001,Rebusco_2005,Rebusco_2006,
Simionescu_2008,Simionescu_2009,Gaspari_2011a,
Gaspari_2012,ettori2013} they are mixed and redistributed throughout the ICM.

Observations at different redshifts of the ICM spatial distribution and
abundance ratios  of different metals  provide important hints on the
number, distribution and metal production efficiency of different SN
types. Indeed, whereas light metals (such as, O, Ne, Mg or Si) are
principally released by core-collapse or Type II supernovae (SNII),
heavier elements (mainly, Fe and Ni) are produced by Type Ia
supernovae (SNIa). On the other hand, while both SNe types contribute
similarly to the production of intermediate-mass elements (e.g., S, Ar or
Ca), SNII and AGB stars contribute to lighter metals, with SNII
providing the bulk of O, Ne, Mg and Si and AGB stars producing the bulk
of C, N and similar elements \citep[e.g.][]{Bohringer_2010,Nomoto_2013,
de_Plaa_2013}.

As a further distinction between the two kinds of SN, while SNII
(stemming from high-mass short-lived stars)  evidence recent star
formation events, SNIa (resulting from long-lived low-mass stars) are
associated to later metal enrichment. Therefore, the analysis of the
ICM metal enrichment and its z-evolution encode crucial information on
the cluster star formation history and on the interplay of a number of
astrophysical processes shaping observed cluster properties.

In the last years, observations of galaxy clusters with X-ray
satellites such as \asca, \sax, \xmm\ or \chandra, have revealed a
number of features of the abundance and distribution of different
metals in local clusters. Early observations already confirmed that
the ICM is characterized by a mean Fe abundance of around $\sim1/3$
the solar value \citep[e.g.][]{Mushotzky_1996,
Finoguenov_2000}. Moreover, the iron abundance distribution within
galaxy clusters has been shown to depend on the cool-coreness of the
considered system: while cool-core (CC) clusters show increasing Fe
abundance radial profiles towards their cores, non-cool core (NCC)
clusters tend to have a flat distribution out to $\sim 0.4 R_{\rm
180}$ \citep[][]{deGrandiMolendi2001, deGrandi_2004, Bohringer_2004,
baldi2007}.  As a consequence, the levels of core ($r<0.1R_{\rm
180}$) entropy and central enrichment are anti-correlated \cite[see,
e.g.,][]{leccardi2010}. However, beyond the central core regions, the
distribution of metals within the ICM is consistent with a flat and
universal radial profile. Besides the abundance of individual
elements, the radial distribution of metal abundance ratios, such as
Si/Fe or O/Fe, also appears to be nearly flat
\citep[see, however,][who found a rising Si/Fe profile above
$\sim0.1-0.2R_{\rm500}$]{Rasmussen_2009}, implying a uniform
distribution and a similar contribution to metals from SNIa and SNII
\citep[e.g.][]{sato2008, sakuma2011, matsushita2013}.

As for the ICM enrichment history, despite the existing observational
difficulties \citep[][]{Molendi_2016}, X-ray observations are generally
consistent with a non-evolving average metallicity distribution out
 to $z\sim1.5$  \citep[e.g.][]{Balestra_2007, Maughan_2008,
 Baldi_2012, ettori2015, McDonald2016,mantz2017}.

Most of the results commented so far refer to relatively central cluster
regions ($r\leq 0.2-0.4 R_{\rm 180}$). Nevertheless, recent X-ray observations
with \suzaku\ of nearby clusters \cite[][]{werner2013, simionescu2015,
urban2017, Ezer_2017}  and groups  \cite[][]{Tholken2016} have provided precise
and well-resolved results on the ICM metallicity distribution out to
cluster outskirts, reaching the virial radius and beyond.  These
observations provide a uniform and universal metal distribution in the
outer regions, suggesting an early ($z\geq2$) enrichment of the ICM~\cite[see also][]{fujita2008}.
As a consequence of these observations, we expect a non-evolving
cluster metal content at $r\geq 0.5 R_{\rm 180}$ and out to high redshift.
These expectations have been recently confirmed by  \cite{mantz2017},
who employed the largest selection of clusters existing so far from X-rays
and SZ surveys within $z<1.2$.
Contrarily to the innermost cluster volume \cite[e.g.][]{dePlaa_2007,
de_Plaa_2013,mernier2016a,mernier2016b}, setting precise constraints
on the ICM metallicity in cluster outskirts still represents a challenge since
these regions are characterized by a  low surface brightness and can be
affected by a significant background contamination and multi-temperature
gas \cite[e.g.][]{Molendi_2016}. Therefore, to explore in more detail the
periphery of clusters it will be necessary the use of next-generation X-ray
missions, such as \athena{}\footnote{\tt http://www.the-athena-x-ray-observatory.eu}
\cite[e.g.][]{athena}, with improved spectral resolutions, larger collecting
area, and a precise characterization of the background.

From the theory side, cosmological hydrodynamical simulations represent
valuable tools to deepen our knowledge of the ICM chemical enrichment
history \citep[see][for a review]{Borgani_2008}. Results from these simulations,
including a number of numerical implementations for the AGN feedback,
the star formation or the chemical enrichment model, are in line with most of the
observational results on the ICM metal content and distribution, its evolution with
redshift and the different metal production efficiencies of SNII and SNIa
\citep[e.g.][]{tornatore2004, tornatore2007, Cora_2008, rasia2008, Fabjan_2008,
Fabjan_2010, mccarthy2010, planelles2014, Martizzi_2016,Biffi_2017,vogelsberger2017}.

In a preliminary analysis, we presented results on the chemical
enrichment of the ICM in
a set of 29 re-simulated galaxy clusters~\cite[][]{Biffi_2017},
selected from cosmological hydrodynamical simulations performed with
an improved version of the SPH code GADGET-3
\cite[][]{springel2005}. These simulations, accounting for the effects
of star formation, chemical enrichment and stellar and AGN feedback,
were able to reproduce the coexistence of CC and NCC clusters in a
simulated volume~\cite[][]{rasia2015}
and both populations showed a good agreement with observed clusters in
terms of entropy and iron abundance profiles, reproducing as well the
central entropy-metallicity anti-correlation.
In \cite{Biffi_2017}, we investigated in detail the ICM metal
distribution for the same sample of clusters.  Specifically, we
explored the abundances of different elements, such as Fe, Si and O,
their spatial and temporal distributions, and the Si/Fe and O/Fe
ratios. We studied the dependencies of these quantities on the
different physical processes included in the simulations and on the
cluster cool-coreness, discussing as well the
different channels for metal production as contributed by SNIa, SNII
and AGB stars. In general, our results were consistent with
observations of low-redshift massive clusters from the core out to the
outskirts.

In the present paper,  we complete the analysis presented in \cite{Biffi_2017}
and we go a step further by investigating in detail the  chemical and dynamical
evolution of the ICM metal-rich gas populating the outskirts of present-day galaxy
clusters. In order to do so, we track back in time the origin of ICM metals
found in cluster  outskirts at $z=0$.
Recalling the importance of iron as a tracer of the ICM metal
  enrichment level in X-ray observations of clusters, the
  chemical properties of the tracked gas will be traced by means of
  its iron abundance.
Being our simulations particle-based, it is also straightforward to follow the
particles selected at $z=0$ and spatially trace them back, recording the
contribution to their metal content due to different enrichment sources
(i.e. SNII, SNIa, AGB stars).

The aim of this analysis is to address a number of questions related to
the connection  between different feedback and enrichment sources, the
origin of the gas located in the  outskirts of clusters at $z=0$ and contributed
by a particular stellar source, or the dynamical evolution of different metal
species depending on their main contributing source.

The paper is organized as follows. In Section \ref{sec:simulations} we briefly
describe the main numerical features of our simulations, the chemical evolution
model employed and the set of clusters to be analyzed.  Section \ref{sec:met}
explores the origin and the spatial and temporal evolution of the metal-rich gas
located in cluster outskirts at $z=0$. The contribution to the metal enrichment
by different sources, such as SNIa, SNII and AGB stars, is also tracked in the
simulations and discussed in Section~\ref{sec:track}.  Finally, in Section
\ref{sec:conclusions}, we discuss and summarize our main results.

\section{Numerical simulations}
\label{sec:simulations}

The following analysis is performed on a representative subsample of
simulated galaxy clusters, analysed also in~\cite{Biffi_2017}.  These
are part of a set of zoom-in re-simulations of 29 Lagrangian regions
extracted from a parent cosmological DM-only simulation, centered on
24 massive clusters ($\mtwo > 8 \times 10^{14}\msunh$) and 5
group-size systems ($ 10^{14} < [\mtwo / \msunh] < 4\times 10^{14}$),
and re-simulated at higher resolution including
baryons~\cite[see][]{bonafede2011}.  The parent cosmological volume is
$1 \,h^{-3}\rm{\,Gpc}^3$ and adopts a $\Lambda$CDM cosmological model
with $\Omega_m=0.24$, $\Omega_b=0.04$, $H_0=100 \times
h=72\kms\mpc^{-1}$, $\sigma_8=0.8$ and $n_s=0.96$ \cite[consistent
  with WMAP-7 constraints;][]{komatsu2011}.

Simulations were performed with a version of the Tree-PM
Smoothed-Particle-Hydrodynamics (SPH) code
GADGET-3~\cite[][]{springel2005}, in which the hydrodynamical scheme
has been modified in order to improve the SPH capabilities of treating
discontinuities and following the development of gas-dynamical
instabilities.  These modifications include higher-order interpolation
kernels and derivative operators as well as advanced formulations for
artificial viscosity and thermal diffusion~\cite[][]{beck2015}.  The
mass resolution of these simulations is $m_{\rm{DM}} = 8.47\times10^8
\, \msunh$ and $m_{\rm{gas}} = 1.53\times10^8\, \msunh$, for DM and
gas\footnote{The gas mass resolution quoted is the intial value. In
  fact, gas masses are allowed to vary during the simulation, as a
  consequence of star formation episodes or accretion of metal mass
  due to chemical feedback from neighbouring star particles.}
particles respectively.  The Plummer-equivalent gravitational
softening length for DM is set to $\epsilon = 3.75\, h^{-1}$\,kpc in physical
units up to $z=2$ and in comoving units at higher redshifts. For gas,
star and black-hole particles it is fixed in comoving coordinates, at
all redshifts, to $3.75\, h^{-1}$\,kpc, $2\, h^{-1}$\,kpc and $2\,
h^{-1}$\,kpc, respectively.

The zoomed-in re-simulations include a large variety of
physical processes describing the evolution of the baryonic component.
These comprise
heating/cooling from Cosmic Microwave Background (CMB) and
from a UV/X--ray time-dependent uniform ionizing
background~\cite[included as in][]{haardt_madau01},
metallicity-dependent radiative cooling~\cite[][]{wiersma_etal09},
star formation~\cite[][]{springel2003} and metal
enrichment~\cite[][]{tornatore2004,tornatore2007} from various stellar
sources.
In the \csf\ version of these runs, it is included only stellar feedback,
in the form of thermal SN feedback and galactic (SN-driven) winds with
a velocity of $350\kms$, according to the original model
by~\cite{springel2003}.
The more complete set of simulations (\agn) additionally includes a
model for gas accretion onto super-massive black holes (SMBH) powering
active-galactic-nuclei (AGN) thermal feedback, presented
in~\citet{steinborn2015} \cite[mimicking the boosting action of the chaotic
cold accretion process; e.g.,][]{Gaspari_2017}.

Results on the thermo- and chemo-dynamical properties of these
simulations have been presented in a series of recent
papers~\cite[][]{rasia2015,villaescusa2016,truong2018,Biffi_2016,Planelles_2017,Biffi_2017},
to which we refer the interested reader for further details.  In
particular, we have shown that the \agn\ runs of the 29 most massive
clusters at the center of the re-simulated Lagrangian regions
reproduce the observed diversity between CC and
NCC populations, with consistent proportions between
the two classes and gas entropy profiles in agreement with
observational data~\cite[][]{rasia2015}.  The central level of entropy
also anti-correlates with the core enrichment
level~\cite[][]{Biffi_2017}, as expected from observational
evidences~\cite[e.g.][]{leccardi2010}, and the ICM thermal pressure,
temperature and metallicity radial profiles, as well as scaling
relations between global observable properties and mass, are in good
agreement with a variety of observational
datasets~\cite[][]{rasia2015,truong2018,Planelles_2017,Biffi_2017}.

We are currently investigating the main properties of the galaxies in
our set of simulated clusters, identified by the SubFind
algorithm~\cite[][]{dolag2009}.  Given the complementarity with
respect to the current analysis on the origin of the ICM chemical
properties, we briefly summarize here some results, while deferring a
detailed study, employing as well higher-resolution simulations, to a
future dedicated work.
  The analysis of the galaxy population in our simulations indicates
  that there is a positive correlation between the total cluster mass
  at $\rfive$ and the stellar mass of the central $30\,\kpc$ of the
  cluster, where the brightest cluster galaxy (BCG) resides, in
  agreement with observational data~\cite[e.g., by][]{kravtsov2014}.
  In addition, the stellar mass and oxygen abundance of the
  star-forming gas in resolved galaxies correlate as expected from
  observations by, e.g.,~\cite{tremonti2004}, \cite{sanchez2013}
  and~\cite{hunt2016}.  Specifically, galaxies with stellar masses
  above $\sim 3\times 10^{10}\msun$ show a similar amplitude and trend
  with respect to the data. Instead, less massive galaxies are in
  broad agreement with the observed datapoints in terms of amplitude,
  given the large scatter between the various datasets, but present a
  steeper relation: galaxies with stellar masses around $\sim
  [1$--$3]\times 10^{10}\msun$ are consistent with data by
  \cite{tremonti2004} while those with stellar masses around $\sim
       [0.5$--$1]\times 10^{10}\msun$ are in better agreement with
       \cite{sanchez2013} and~\cite{hunt2016}.

  Similar to recent X-ray observations~\cite[][]{mantz2017}, our
  simulations also present a metallicity-mass anti-correlation within
  $0.1\,\rfive$ roughly corresponding to the location of the BCG,
  whereas on global scales (i.e. within $\rfive$) the ICM metallicity
  is found to vary very little with the system mass, from group to
  cluster scales (see Truong et al., {\it in prep.}).

\smallskip

{\it Chemical enrichment model.}  The chemical evolution model
implemented in our simulations follows the formulation presented in
detail in~\cite{tornatore2004,tornatore2007}.

  We only recall here the main features of this model.
Heavy elements are produced through three main enrichment
channels, namely by SNIa, SNII and AGB stars, depending on their typical
life-times and yields.
To this purpose, we assume the initial mass function
  by~\cite{chabrier03} as a distribution of initial masses for the
  population of stars, and the mass-dependent life times
  by~\cite{padovani_matteucci93} to account for the different time
  scales of stars of different masses.
The abundance of the various metal species produced during the
evolution of a stellar particle is then estimated by considering
different sets of stellar yields for SNIa, SNII and AGB
(namely~\citealt{Thielemann2003}, for SNIa;
\citealt{WoosleyWeaver1995} combined with those
by~\citealt{romano2010}, for SNII; \citealt{karakas2010}, for AGB
stars).
Once metals are produced, they are distributed from stars to
  surrounding gas particles by smoothing them on the SPH kernel,
  consistently with the other thermodynamical quantities, while no
  process of metal diffusion is explicitly included in the
  simulations.
Previous studies in the literature have investigated the influence of
changing the details of the metal spreading on the final enrichment
pattern in simulated
clusters~\cite[e.g.][]{tornatore2007,wiersma_etal09}.  From various
tests presented in~\citealt{tornatore2007} for the same chemical
evolution model used in the current simulations, we conclude that
changes in the numerical details of the metal injection scheme
(e.g.\ number of neighbouring particles or weighting scheme for the
metal diffusion to the gas) have only a minor impact on the resulting
star formation history, stellar population and chemical enrichment
pattern.  The differences, in particular, are smaller than the scatter
from cluster to cluster.  Moreover, for a given scheme of metal
spreading, the final enrichment level of the simulated clusters is
more strongly influenced by the physical processes included in the
smulations, such as the energy
feedback~\cite[see][]{Fabjan_2010,planelles2014,Biffi_2017}.

We follow the production and evolution of 15 chemical
species: H, He, C, Ca, O, N, Ne, Mg, S, Si, Fe, Na, Al, Ar and Ni.
For each gas particle, we can trace not only the chemical composition,
but also the fraction of each metal which is produced by the three
enrichment sources (SNIa, SNII, AGB).

In order to investigate the chemical enrichment of the ICM in the
simulated galaxy clusters we explicitely compute the mass fractions of all
the metals in the gas --- also distinguishing among the enrichment
sources.  We will also study the iron abundances, since iron is typically used as
metallicity tracer in cluster X-ray observations. For this, we will
provide iron abundances in units of its solar value ($4.68\times
10^{-5}$, iron number fraction relative to hydrogen) according
to~\cite{angr1989} (hereafter ANGR89).

\smallskip

{\it Dataset.}  The following analysis has been performed on the
representative subsample of 4 clusters~\cite[already presented
  in][]{Biffi_2017}, chosen to include two massive and two smaller
systems, in both cases comprising one CC and one NCC cluster.
Results will be shown and discussed in particular for one
study-case cluster (D2),  that is a small CC object. Similar results were found for all
the four systems, essentially with no dependence on either mass or
cool-coreness (see discussion in Appendix~\ref{app:others}).
\begin{table}
  \begin{tabular}{lcccc}
  \hline
  cluster & $\mvir$ & $\rvir$ & $\rtwo$ & cool-\\
   &  [$10^{14}\msunh$] & [$\mpch$] & [$\mpch$] & coreness\\
  \hline
  D2  & 5.26  & 1.70 & 1.19 & CC  \\
  D3  & 6.53  & 1.83 & 1.28 & NCC \\
  D6  & 15.41 & 2.43 & 1.69 & NCC \\
  D10 & 15.46 & 2.43 & 1.65 & CC \\
\hline
\end{tabular}
\caption{Main properties of the four systems analysed, at $z=0$. From
  left to right, the columns report the cluster id, viral mass
  ($\mvir$), virial ($\rvir$) and $\rtwo$ radii, and the CC/NCC
  classification based on the core thermal
  properties.}\label{tab:glob-prop}
\end{table}
The global properties of the four clusters, namely virial mass and
radius, $\rtwo$ radius\footnote{The cluster virial radius, $\rvir$ ($\rtwo$),
  is defined as the radius enclosing an average overdensity
  $\Delta_{\rm vir}$ ($200$) times the critical cosmic density,
  $\rho_c(z)=3H(z)/(8\pi G)$, as predicted by the spherical collapse
  model \citep[e.g.][]{bryan1998}. The virial mass ($\mvir$),
  correspondingly, is the mass enclosed within $\rvir$.} and cool-core
or non-cool-core \cite[CC/NCC; see][for details on the
classification criteria]{rasia2015} classification are listed in
Table~\ref{tab:glob-prop}, for the reference \agn\ simulation set.

\section{Metal enrichment of cluster outskirts}\label{sec:met}

\begin{figure*}
  \includegraphics[width=0.7\textwidth]{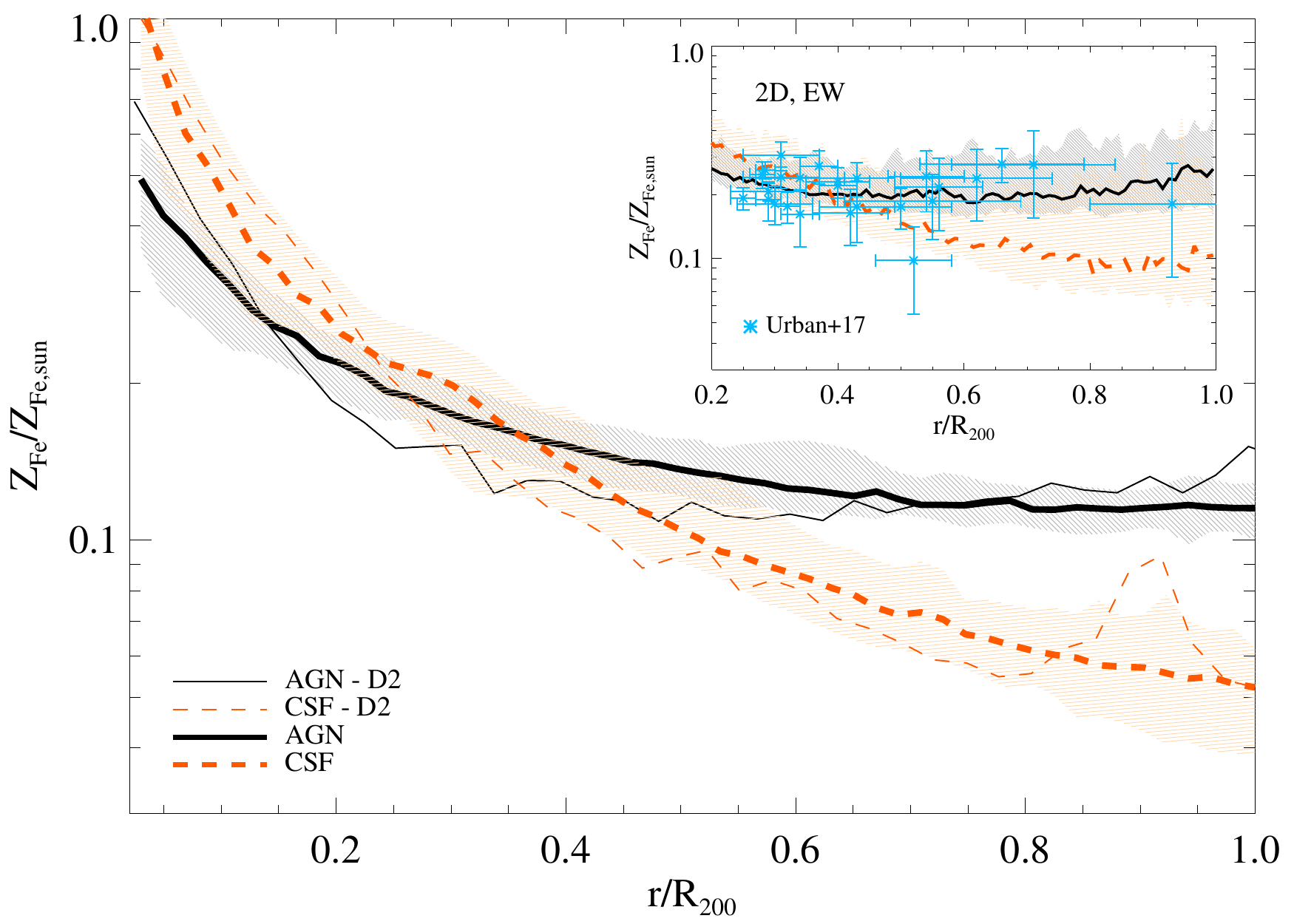}
  \caption{Comparison between the median Fe-abundance profiles
    (w.r.t.\  solar values by~ANGR89) of the 29 most massive haloes
    in our set of simulated regions in the two simulation runs with
    (`\agn') and without (`\csf') the treatment for AGN
    feedback. Thick lines stand for the median profiles of the whole
    simulated sample, whereas thin lines represent the corresponding
    profiles of the region D2. These Fe-abundance profiles are
    mass-weighted and three-dimensional. The shaded areas around the
    median profiles represent the $16^{\rm th}$ and $84^{\rm th}$
    percentiles of the distributions.  {\it Top-right inset:}
    Comparison between the emission-weighted, two-dimensional median
    Fe-abundance profiles of our sample of 29 most massive haloes in
    the \agn\ and \csf\ simulations (see main legend) and the
    abundance measurements in the outskirts ($r > 0.25\,\rtwo$) of a
    sample of nearby clusters by~\protect\citet{urban2017} (blue asterisks
    with error bars).}
\label{fig:profs-fe-comp}
\end{figure*}

The metal enrichment of the ICM in cluster outskirts is observed to be
relatively homogeneous, as shown by the flattening of the metal
abundance radial profiles at large cluster-centric distances
\cite[e.g.][]{Leccardi_2008, werner2013, simionescu2015, urban2017,
  mernier2017}.  This is also found in hydrodynamical
simulations of galaxy clusters, when the feedback from AGN is
included~\cite[see][]{Biffi_2017}.  In \figref{fig:profs-fe-comp},
we show the comparison between the median Fe-abundance profiles for our
sample of 29 simulated clusters, namely the most massive ones in each
of the re-simulated Lagrangian regions at $z=0$, for the runs with
(\agn; thick black solid line) and without (\csf; thick red dashed line)
the inclusion of AGN feedback.
These iron profiles are mass-weighted (MW) and three-dimensional, and the shaded areas
in the figure mark the dispersion around the median profile, using the
$16^{\rm th}$ and $84^{\rm th}$ percentiles of the distribution.
From the comparison, we note that in the outer regions the median Fe
profile in the \agn\ case is flatter and higher (by a factor of $\sim
1.6$--$2.2$ in the range $\sim 0.7$--$1\rtwo$) than in the
\csf\ one~\cite[consistently with results presented in][]{Biffi_2017},
with minimal scatter across the sample in both sets of simulations.
The flatness of the Fe-abundance profiles in the outer regions, shown
in the \agn\ clusters, is in agreement with observational
findings~\cite[see also results in][]{Biffi_2017}. In fact, the inset
in \figref{fig:profs-fe-comp} shows a comparison between the median
Fe-abundance profiles of our sample of 29 clusters in the \agn\ (thick
black solid line) and \csf\ (thick red dashed line) simulations and
the metal abundance measurements of a sample of nearby clusters by
\citet{urban2017} (blue asterisks with error bars --- here values from
their Table~5 have been rescaled with respect to the solar reference
value by ANGR89, for consistency with the rest of our paper).  The
radial range shown is a zoom onto the intermediate and outer regions,
i.e.\ $r > 0.25\,\rtwo$, where the observational data
by~\cite{urban2017} indicate a relatively flat Fe-abundance level
($Z_{\rm Fe}=0.316\pm 0.012$ solar, w.r.t. the reference value
by~\cite{asplund2009}, which correspond to $Z_{\rm Fe}\sim 0.21$
when the solar abundances by ANGR89 are considered, as it is in our case)
across the observed clusters. In order to
perform a more faithful comparison \citep[see discussion
  in][]{Biffi_2017}, we show here the emission-weighted (EW)
 projected median Fe-abundance profiles for our simulated set of
clusters. The shaded areas around the median profiles mark the
$16^{\rm th}$ and $84^{\rm th}$ percentiles of the distributions.
  At large cluster-centric distances the scatter across the sample is
  larger than in the case of the MW, 3D profiles. This
    is essentially driven by the presence of substructures, more
    numerous in a larger volume and more impacting in the \csf\ runs
    for the higher emissivity of these clumps. As a net consequence,
    the two sets of profiles result to be marginally consistent
  within 1-$\sigma$. Despite this, we still note that the median
  trends are very different: significantly flatter and higher for the
  \agn\ clusters and steeper and lower for the \csf\ ones.  In this
  respect, the \agn\ median profile shows a remarkable agreement with
  the observational data points, in particular for
  $r>0.5$--$0.6\,\rtwo$.
%

By means of a representative study case from the \agn\ sample -- D2,
whose Fe-abundance profiles are marked with the thin lines in
\figref{fig:profs-fe-comp} -- we investigate in this section the
details and origin of the ICM chemical enrichment in the cluster
outskirts at $z=0$ and the role of AGN feedback in establishing this
pattern.

Specifically, we select the hot-phase gas residing in the external
region of the cluster, within the spherical shell comprised between
$0.75\rtwo$ and $\rtwo$, at the present time $z=0$.
In order to select the hot-phase ICM at $z=0$, we exclude from our
computation those gas particles that have temperature below $3\times
10^4\,$K or a cold-gas component greater than $10$ per
cent. Additionally, we exclude explicitly the gas that is in the
star-forming phase, i.e. that with density greater than the
star-formation density threshold adopted in our code $\rho_{\rm sf\_th} =
0.1$\,cm$^{-3}$.
  In the following, this gas -- or a subselection of it depending on
  its $z=0$ iron abundance -- will be tracked back in time
  in order to investigate its chemical properties at higher redshifts,
  its origin and spatial distribution. For such analysis no
  restriction on the thermal phase of the tracked gas at redshift
  $z>0$ will be made (and MW abundances will be always weighted by the
  total gas mass instead of the hot-phase mass only, for this
  reason).
  Averages will be computed considering the MW value,
  with the purpose of exploring the intrinsic enrichment properties of
  the tracked gas and not of comparing X-ray-like estimates against
  observed data.
  In the following, the main halo at any redshift $z>0$ is
  the main progenitor of the cluster considered at $z=0$.

  With this analysis we intend to explore
  {\it directly} the details of the chemical and dynamical evolution
  of the ICM in the periphery of present-day
  clusters.  In particular, this will help to answer some key
  questions like: where does the metal-rich gas in the outskirts of
  local clusters come from?  Is the  chemical enrichment of
  the ICM in these outer regions mostly happening in situ
  or does it require some
  pre-enrichment at early epochs?  How did different enrichment
  sources, like SNIa and SNII, contribute to the final enrichment
  pattern?

\subsection{Origin of the gas in the cluster outskirts at $z=0$}

\begin{figure}
  \includegraphics[width=0.45\textwidth,trim=10 0 20 24,clip]{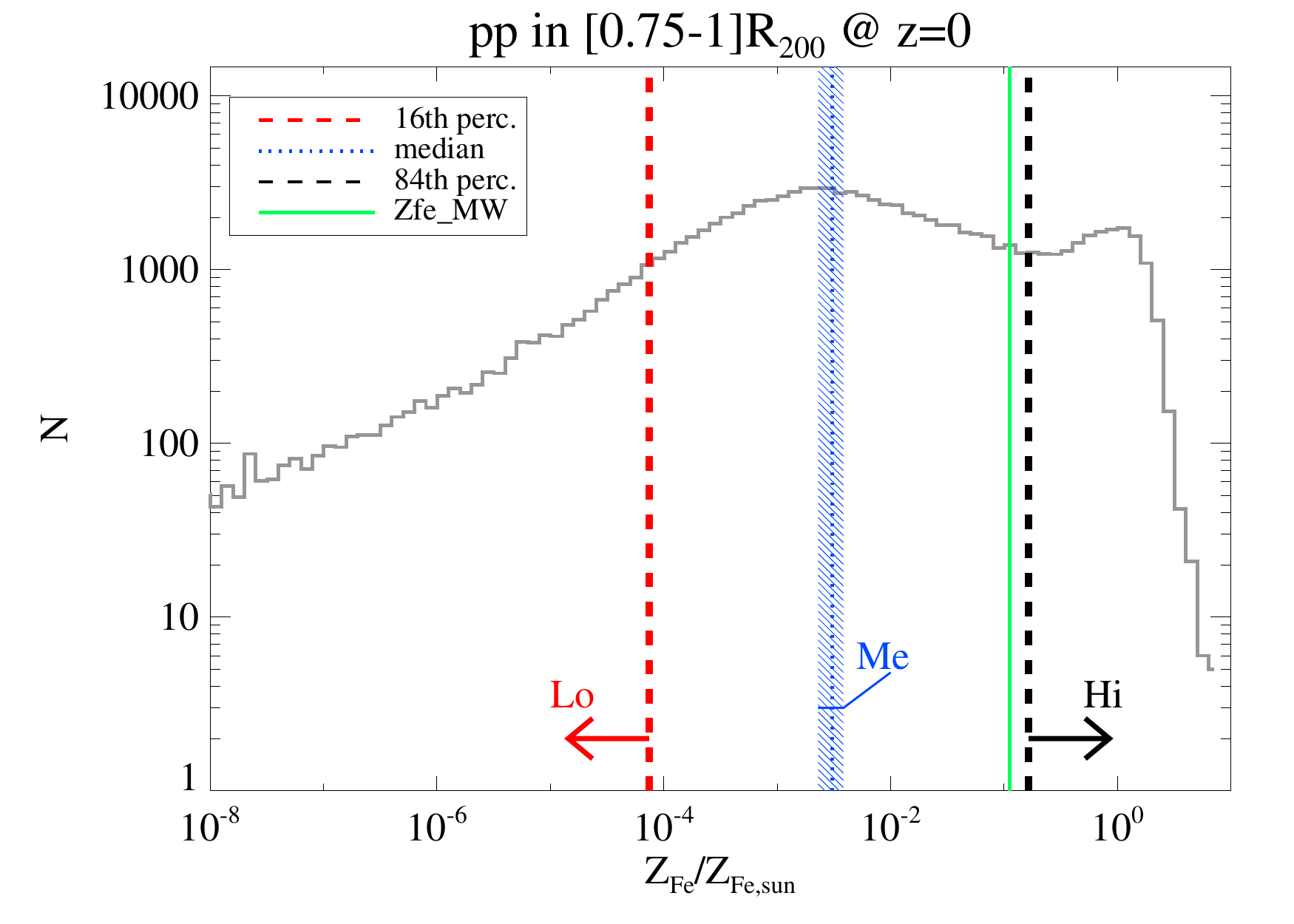}
  \caption{Distribution of the Fe abundance (w.r.t.\ solar values
    by~ANGR89) for the hot-phase gas particles selected to reside
    within $[0.75$--$1]\rtwo$ at $z=0$. As specified in the legend,
    vertical lines mark: the $16^{th}$ percentile (red, dashed), the
    median value (blue, dotted -- with the shaded area indicating
    $\pm 25$\% from this value), the MW value (green, solid line)
    and the $84^{th}$ percentile (black,
    dashed) of the $Z_{\rm Fe}$ distribution. In this
    figure we also indicate the three gas subselections considered in
    Section~\ref{sec:3subsel}.}\label{fig:distrib}
\end{figure}

The hot, diffuse gas selected at $z=0$ in the outer region of the D2
cluster, comprised between $0.75\rtwo$ and $\rtwo$
is characterised by a large
distribution, as it can be seen from \figref{fig:distrib}, where
the median $Z_{\rm Fe}$ value (blue, dotted line -- with the shaded area
indicating $\pm 25$\% from this value), together with the $16^{th}$
(red, dashed line) and the $84^{th}$ (black, dashed line) percentiles
are also reported.
From this distribution, we notice that the gas in the
  outskirts of our simulated cluster is skewed towards high iron
  abundance values, with a low-$Z_{\rm Fe}$ tail and two main peaks, a
  broad one around $\sim 10^{-3}$--$10^{-2}\,Z_{\rm Fe,\odot}$ and a second peak around
  solar values.
  This gas has a typical MW
  iron abundance of about $0.1 Z_{\rm Fe,\odot}$, consistent with the
  profiles shown in \figref{fig:profs-fe-comp}.
  We note that the value of the MW Fe abundance,
  marked with a green solid line, is very close to the $84^{th}$
  percentile of the distribution.  This means that most of the mass of
  the hot gas component is associated to the high-metallicity component of
  the distribution.
A deeper discussion of the distribution shown in
  \figref{fig:distrib} is presented in Section~\ref{sec:distrib-evol}.

In the following, we study the spatial and chemical
  evolution of this hot gas located in the outskirts of the
  main halo of D2 at $z=0$.
  \begin{figure*}
  \includegraphics[width=0.9\textwidth]{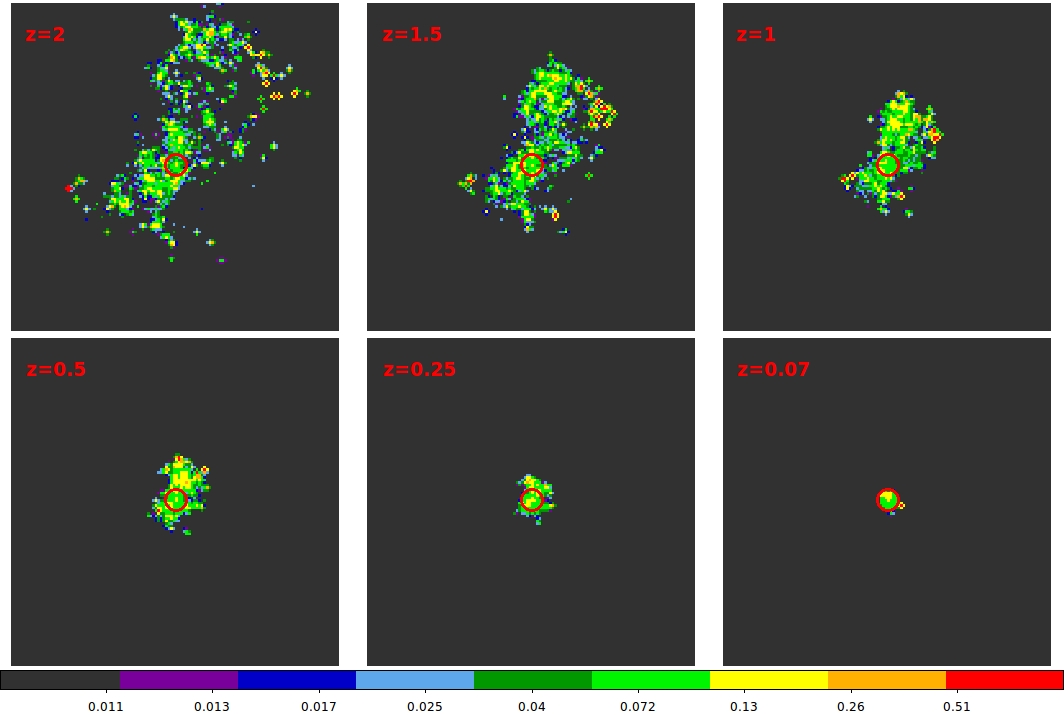}\\
$Z_{\rm Fe}/Z_{\rm Fe,\odot}$
  \caption{Maps of the spatial distribution, at various redshifts
    between $z=2$ and $z=0.07$, of the gas residing in the outer regions -- i.e.\ within
    $[0.75$--$1]\,\rtwo$ -- of D2 at $z=0$. The color bar stands for the MW iron abundance, $Z_{\rm MW,Fe}$
    of the selected particles.
    The red circle represents the virial radius of the main halo, $\rvir$, at any given redshift and each map is $30\rvir$ per side.
  }\label{fig:maps0}
  \end{figure*}%
  To this purpose, we track the selected gas back in time up to
  redshift $z=2$, and analyse its spatial distribution and metallicity
  (primarily using the Fe abundance, $Z_{\rm Fe}$). The map of MW iron
  abundance ($Z_{\rm MW,Fe}$), for the tracked gas is shown in
  \figref{fig:maps0}.  Qualitatively, particles within the cluster
  outskirts at $z=0$ are located, at higher redshifts, quite far from
  their current positions, namely at distances corresponding to
  several times the virial radius of the main cluster progenitor
  (marked in each map).  The observed spatial evolution is the
  result of a highly dynamical process, where the merging of
  substructures and the gas accretion onto the main central halo
  determine the final enrichment pattern.

\begin{figure}
  \includegraphics[width=0.45\textwidth,trim=10 0 20 22,clip]{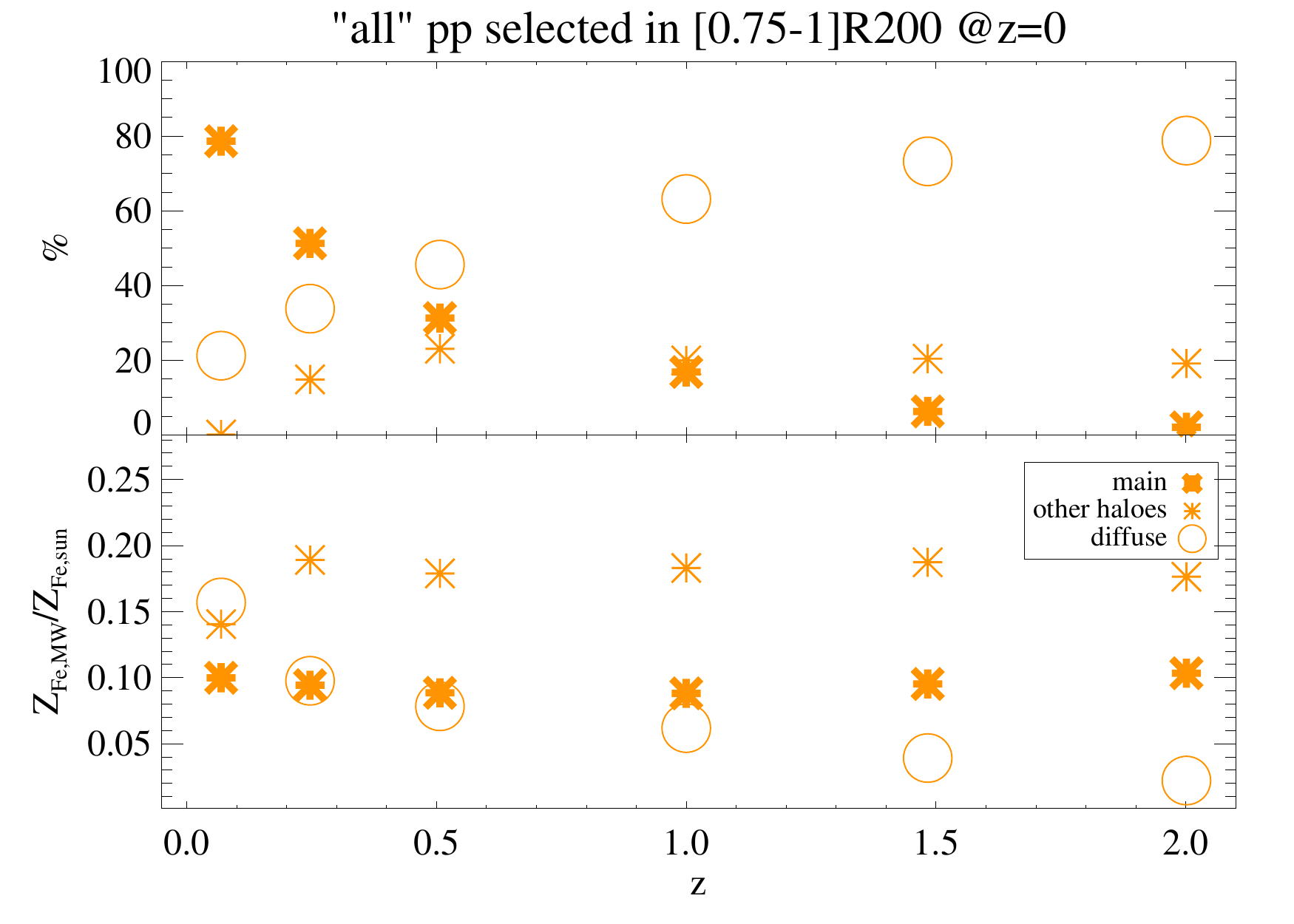}
  \caption{{\it Top:} percentage of the tracked particles that reside
    inside the virial radius of the main-halo progenitor, inside the
    virial radius of any other halo in the re-simulated region with
    mass $\mvir > 10^{11}\msunh$, or is rather outside of them --
    marked by different symbols/sizes, as in the legend -- at various
    redshifts $0.07 <z<2$.  {\it Bottom:} MW Fe abundance (in solar
    units by~ANGR89), $Z_{\rm MW,Fe}$, for the same selection used in the
    upper inset.%
    \label{fig:pp_inRvir_all}}
\end{figure}

A more quantitative description of this evolution is provided in
\figref{fig:pp_inRvir_all}, where we show the percentage of the {\it
  tracked gas} that resides, at a given redshift $z$, inside the
virial radius of the main halo (thick asterisks), within the virial
radius of any surrounding halo more massive than
$10^{11}\msunh$ (thin asterisks) or outside of
them all (empty circles), either in very small haloes or in the
diffuse component.
  The chosen mass threshold of $10^{11}\msunh$ allows us to restrict
  only to haloes that are resolved with a reasonable number of
  particles.
  Given our current resolution, we merely consider the haloes and
  their virial boundary to locate the origin of the tracked gas, while
  a detailed analysis of these haloes is not allowed, especially at
  high redshift.

In the lower inset of \figref{fig:pp_inRvir_all} we also show the corresponding
MW iron abundance, $Z_{\rm MW,Fe}$.
   From the analysis of this figure we infer several interesting
   insights. At $z=0.5$, more than 50\% of all the tracked gas
   particles is still outside of any halo with $\mvir >
   10^{11}\msunh$, included the main one, and is rather enclosed in
   small galaxies or in the form of diffuse gas.  This percentage
   significantly increases up to $\sim 80\%$ at redshift $z=2$.
   Essentially, we see that for this cluster the peripheral gas has
   been accreted primarily from smaller haloes and from the diffuse
   component in the redshift range $0.5$--$2$, when in fact the
   percentage enclosed within intermediate-mass neighbouring haloes
   remains roughly constant.

   Over this whole redshift range, the MW Fe-abundance of the tracked
   gas that is found within the virial radius of the main progenitor
   at each redshift $z$ does not change significantly.
   This balance is reached through a combination of different phenomena:
   the continuous accretion of metal-poor gas,
   the accretion of gas that has been significantly enriched at higher
   redshift and is either diffuse or in very small sub-haloes that
   merge onto the main cluster, and some residual enrichment from
   low-redshift star formation or from long-living stars.
   \begin{figure*}
  \includegraphics[width=0.85\textwidth]{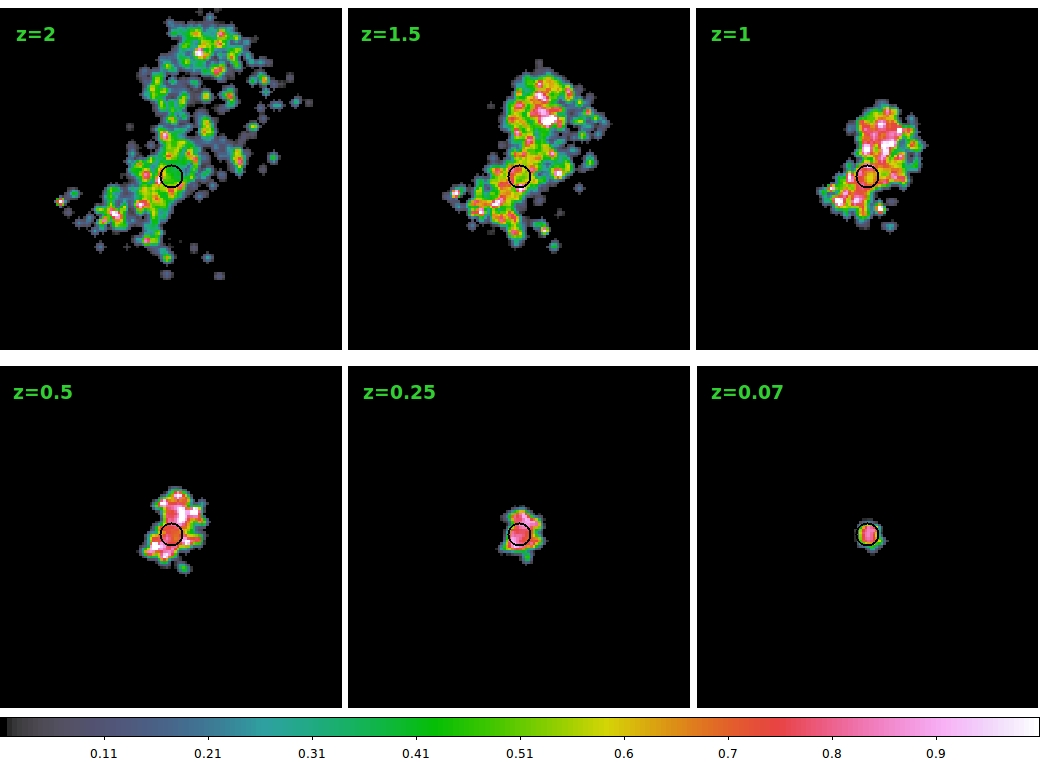}\\
$Z_{\rm Fe}/Z_{\rm Fe,\odot}$
  \caption{Maps of the spatial distribution, at various redshifts
    between $z=2$ and $z=0.07$, of the high-$Z_{\rm Fe}$ gas (\hi)  residing in the outer regions -- i.e.\ within
    $[0.75$--$1]\,\rtwo$ -- of D2 at $z=0$. The color bar stands for the MW iron abundance, $Z_{\rm MW,Fe}$
    of the selected particles.
    The red circle represents the virial radius of the main halo, $\rvir$, at any given redshift and each map is $30\rvir$ per side.
      \label{fig:maps}}
   \end{figure*}

   Similarly, the MW Fe abundance of the tracked gas that is enclosed
   within the surrounding haloes is also relatively constant from
   $z=2$ down to $z=0.25$, despite a higher normalization with respect
   to the main cluster. We verified and found that the star formation
   rate (per unit gas mass) for this gas, locked in the surrounding
   intermediate-mass haloes, is typically higher than for the gas in
   the main cluster progenitor suggesting that star formation is still
   ongoing and continuous metal production is consequently more
   significant. Despite the fact that we are considering here only the
   gas selected at $z=0$ in the main cluster periphery and tracked
   back in time, this is consistent with the expectation that
   lower-mass systems show a slightly higher content of metals, in
   mass, with respect to the relative total gas mass~\cite[see
     e.g.][]{yates2017}.
   Instead, a mild evolution of the MW Fe-abundance of the gas
   outside of any halo more massive than $10^{11}\msunh$ (including
   the main) is observed, as indeed it increases in the range
   $\sim 0.02$--$0.15\,Z_{\rm Fe,\odot}$, from $z=2$ to $z=0.07$.
   This can be explained by the fact that the large amount of diffuse
   gas at high redshift must be still pristine or at least very poorly
   enriched, partly diluting the average $Z_{\rm MW,Fe}$.  As redshift
   decreases, residual star formation and enrichment from SNIa and AGB
   contribute to pollute the gas and gas depletion due to the
   accretion onto the main cluster will combine to produce a net increase
   of the average iron abundance of the gas that remains in the diffuse
   component.

   Nonetheless, we conclude that also the diffuse gas has a typical
   abundance that is overall similar to the one already within the
   cluster, and the other haloes, so that the evolution of the
   outskirts metallicity below $z<2$ is not expected to be
   significant. The gas, diffuse and within merging haloes, that is
   continuously accreted by the forming cluster, has in fact a typical
   iron abundance which varies only mildly with redshift.
   This picture is also consistent with recent numerical results from
   cosmological hydrodynamical simulations performed with the
   moving-mesh code Arepo, as presented in~\cite{vogelsberger2017}.

\begin{figure*}
\centering
\includegraphics[width=0.87\textwidth]{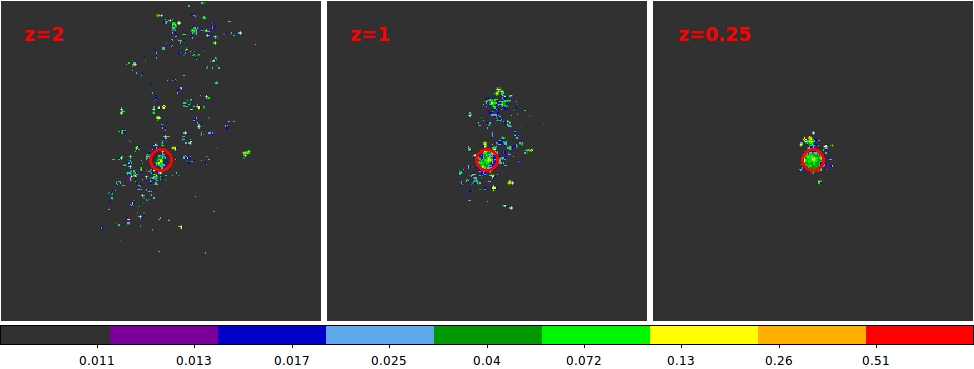}\\
$Z_{\rm Fe}/Z_{\rm Fe,\odot}$
  \caption{Similar to \protect\figref{fig:maps0}, but for the tracked gas in the CSF simulation of D2.
  Here only three redshifts are shown ($z=2,1,0.25$).\label{fig:maps0_CSF}}
  \includegraphics[width=0.87\textwidth]{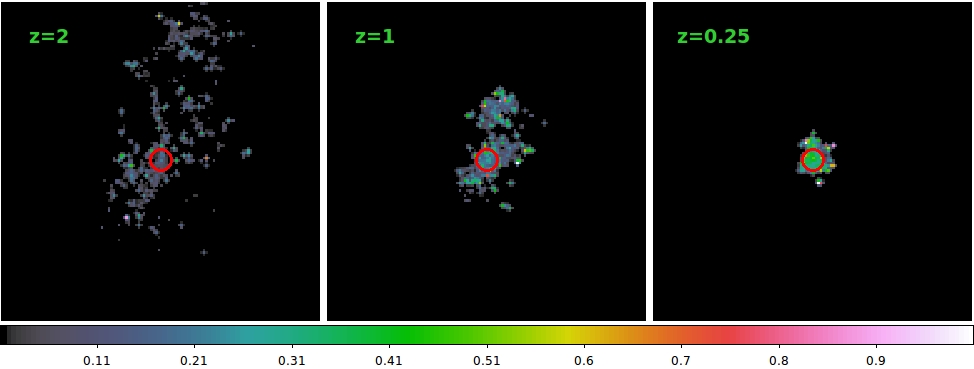}\\
$Z_{\rm Fe}/Z_{\rm Fe,\odot}$
  \caption{Similar to \protect\figref{fig:maps}, but for the tracked \hi\ gas component in the CSF
  simulation of D2. Here only three redshifts are shown ($z=2,1,0.25$).\label{fig:maps_CSF}}
\end{figure*}

\begin{figure}
  \centering
  \includegraphics[width=0.45\textwidth,trim=10 0 15 22,clip]{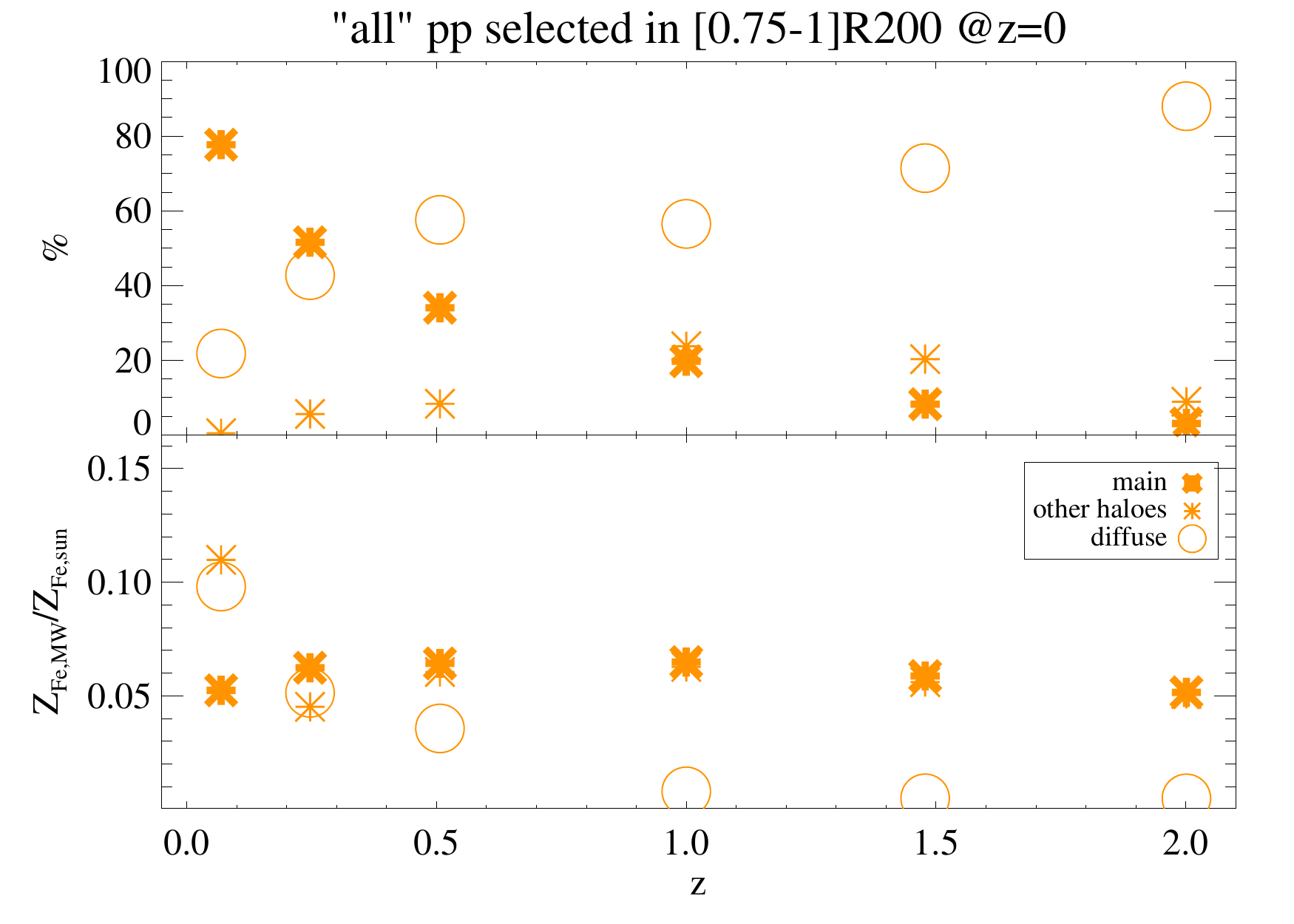}
  \caption{Similar to \protect\figref{fig:pp_inRvir_all} but for the
    CSF simulation of D2.\label{fig:pp_inRvir_Fe_CSF}}
\end{figure}

   By inspecting the \csf\ simulation of the same cluster, instead, we
   find that the ICM in the present-day outskirts presents a different
   history. Although accreted as well from the surrounding
   proto-cluster region, the majority of the tracked gas has lower
   metal content and the metal-rich gas is remarkably clumpier,
   especially at high redshifts.
   This can be inferred from the comparison between \figref{fig:maps0_CSF},
   done for the \csf\ simulation of D2, and \figref{fig:maps0}.
   The MW iron abundance of the
   tracked gas is significantly lower at all redshifts up to $z=2$,
   being on average around $0.05\,Z_{\rm Fe,\odot}$ for the gas enclosed
   within the haloes, main or neighbour ones, and about one order of
   magnitude lower for the gas outside (see
   \figref{fig:pp_inRvir_Fe_CSF}). This is consistent with the idea
   that the enriched gas is mostly confined to SF regions when no AGN
   feedback is included, since stellar feedback alone is not enough to
   displace the gas before it is converted into new generation of
   stars by the very efficient, un-quenched star formation. This
   difference at high-redshifts is then propagated into lower-redshift
   profiles and metal distributions, especially at large distances
   from halo cores.

\subsection{Spatial evolution of metal-rich gas in the outskirts}
\label{sec:3subsel}

Considering the level of ICM chemical enrichment in cluster outskirts,
which is more homogeneous and higher when both stellar and AGN
feedback are included in the simulations, it is interesting to
investigate the origin of this metal content: {\it where does the gas
  that is highly-enriched at $z=0$ come from, and what was its
  metallicity at higher redshift?}

Given the typical distribution of iron abundance of the selected gas
particles at $z=0$, reported in \figref{fig:distrib}, we want now to
compare the enrichment history of the gas showing extreme enrichment
levels at z=0 with the average behavior. To this purpose, we select 3 subsets of gas particles
to trace back in time and space.
Namely, we consider:
\begin{itemize}
\item \hi: the highest-$Z_{\rm Fe}$ gas at $z=0$, i.e.\ with
  $Z_{\rm Fe}>Z_{\rm Fe}^{84th}$
\item \me: the gas with median $Z_{\rm Fe}$ ($\pm 25\%$) at $z=0$
\item \lo: the lowest-$Z_{\rm Fe}$ gas at $z=0$, i.e.\ with
  $Z_{\rm Fe}<Z_{\rm Fe}^{16th}$
\end{itemize}
where $Z_{\rm Fe}^{84th}$ and $Z_{\rm Fe}^{16th}$ are the values corresponding
to the $84^{th}$ and $16^{th}$ percentile of the $Z_{\rm Fe}$ distribution,
respectively (the 3 subselections are also marked in \figref{fig:distrib}).
In a similar way to what is done in \figref{fig:maps0}, we can
specifically track the highly enriched gas selected in the cluster
outskirts at $z=0$ (\hi\ gas particles) back in time up to
$z=2$.  In this way, we explore its spatial distribution with respect
to the main progenitor and the surrounding haloes in the region, and
its metallicity at higher redshift. The map of the spatial
distribution and the MW iron abundance for the \hi\ tracked gas is
shown in \figref{fig:maps}.  In general, we note that, even at high
redshift, there are high-$Z_{\rm Fe}$ peaks far beyond the boundary of the
main halo, whose virial radius at
any corresponding redshift is delimited by the black circle.  The
global similarity between these maps and those in \figref{fig:maps0}
indicates that also the high-$Z_{\rm Fe}$ gas residing in the cluster
outskirts at $z=0$ has been accreted from the large-scale
structure around the main halo, during the collapse and accretion of
neighbouring haloes.
  Peaks of iron
  abundance are present already at $z\gtrsim1$ and located in the map
  far away from the central progenitor (at distances up to $\sim 10\rvir$).

\begin{figure}
  \centering
  \subfigure[\hi\ gas.\label{fig:pp_inRvir_Fe-hi}]{\includegraphics[width=0.45\textwidth,trim=15 0 15 22,clip]{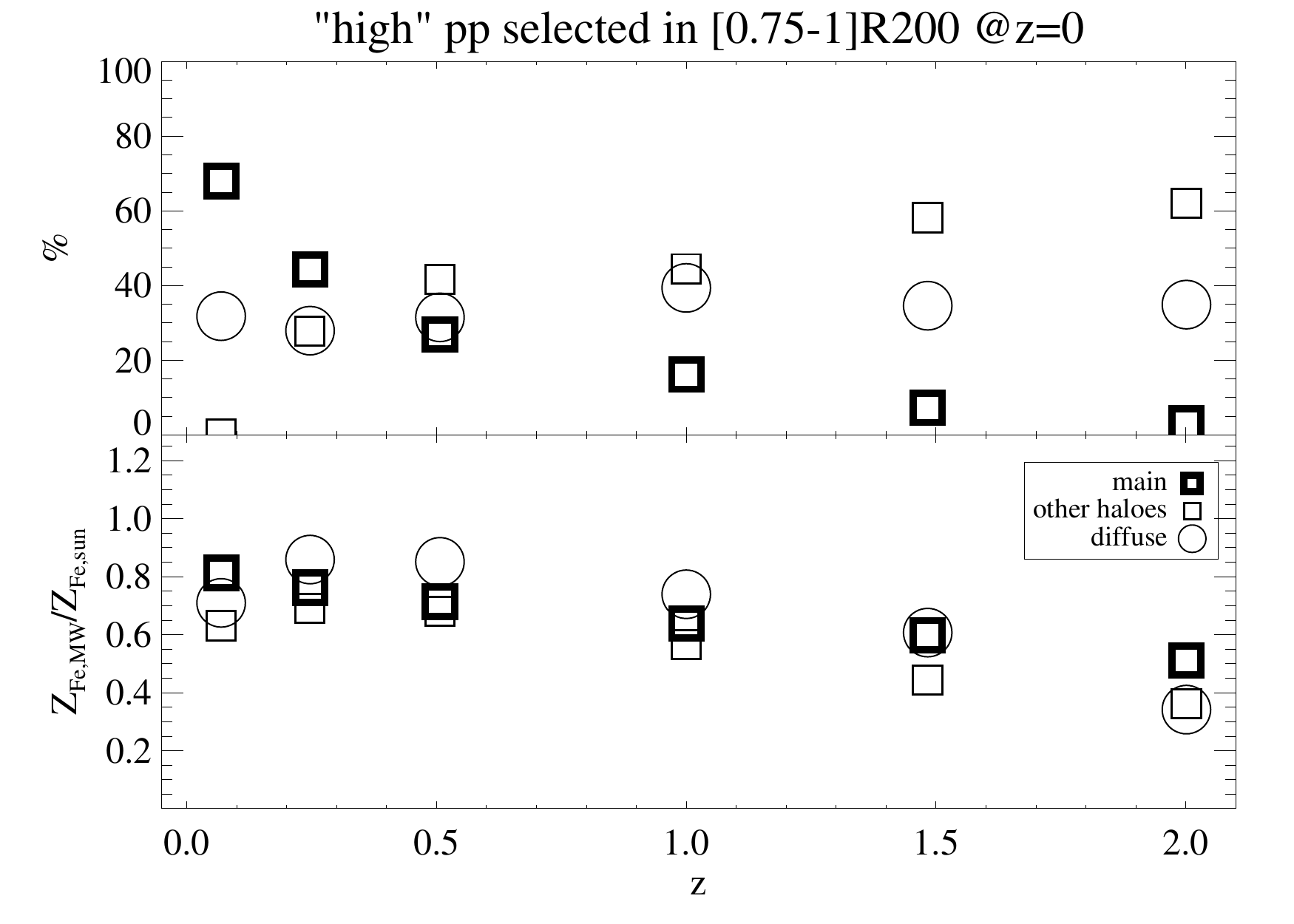}}
  \subfigure[\me\ gas.\label{fig:pp_inRvir_Fe-med}]{\includegraphics[width=0.45\textwidth,trim=15 0 15 22,clip]{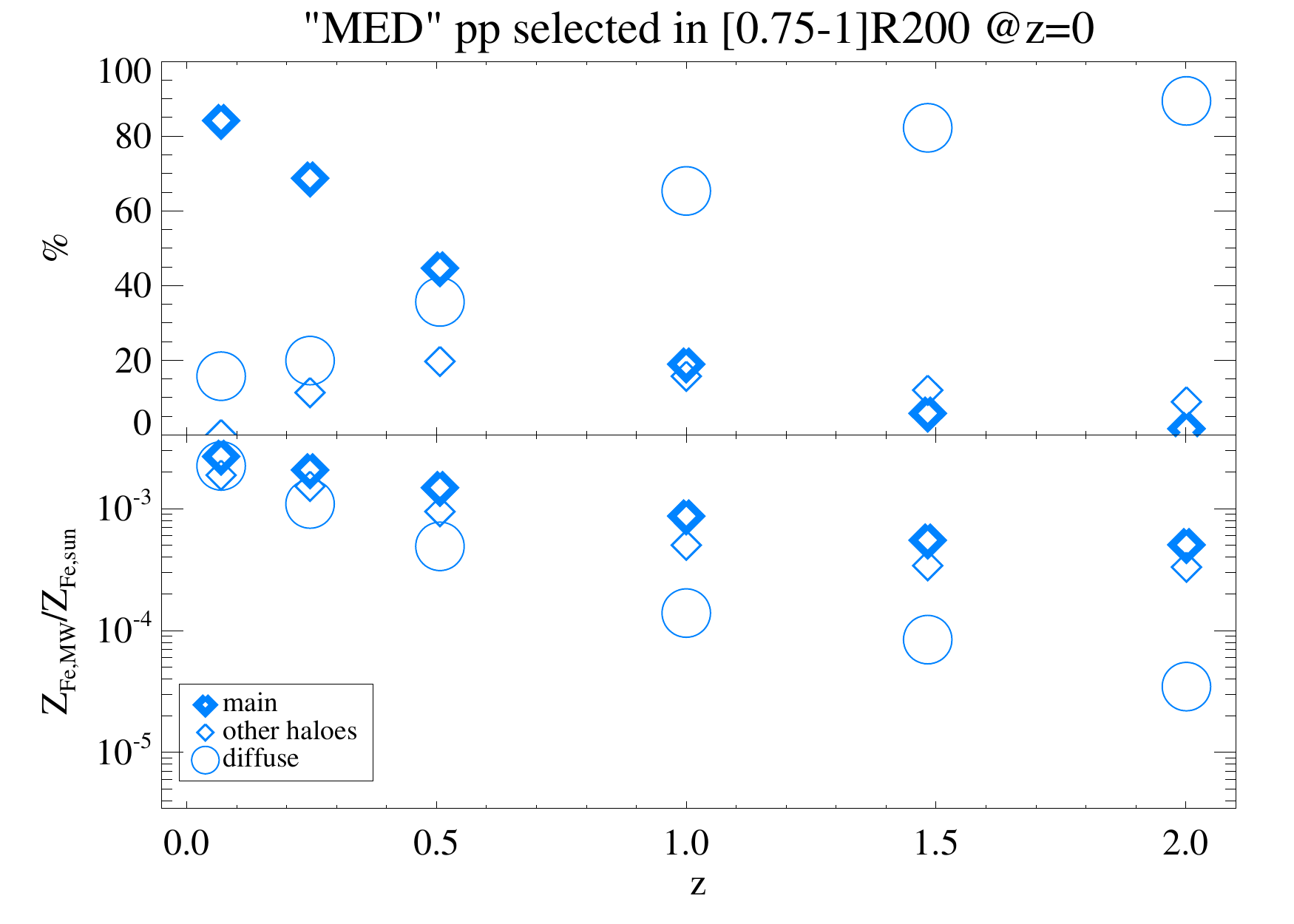}}
  \subfigure[\lo\ gas.\label{fig:pp_inRvir_Fe-lo}]{\includegraphics[width=0.45\textwidth,trim=15 0 15 22,clip]{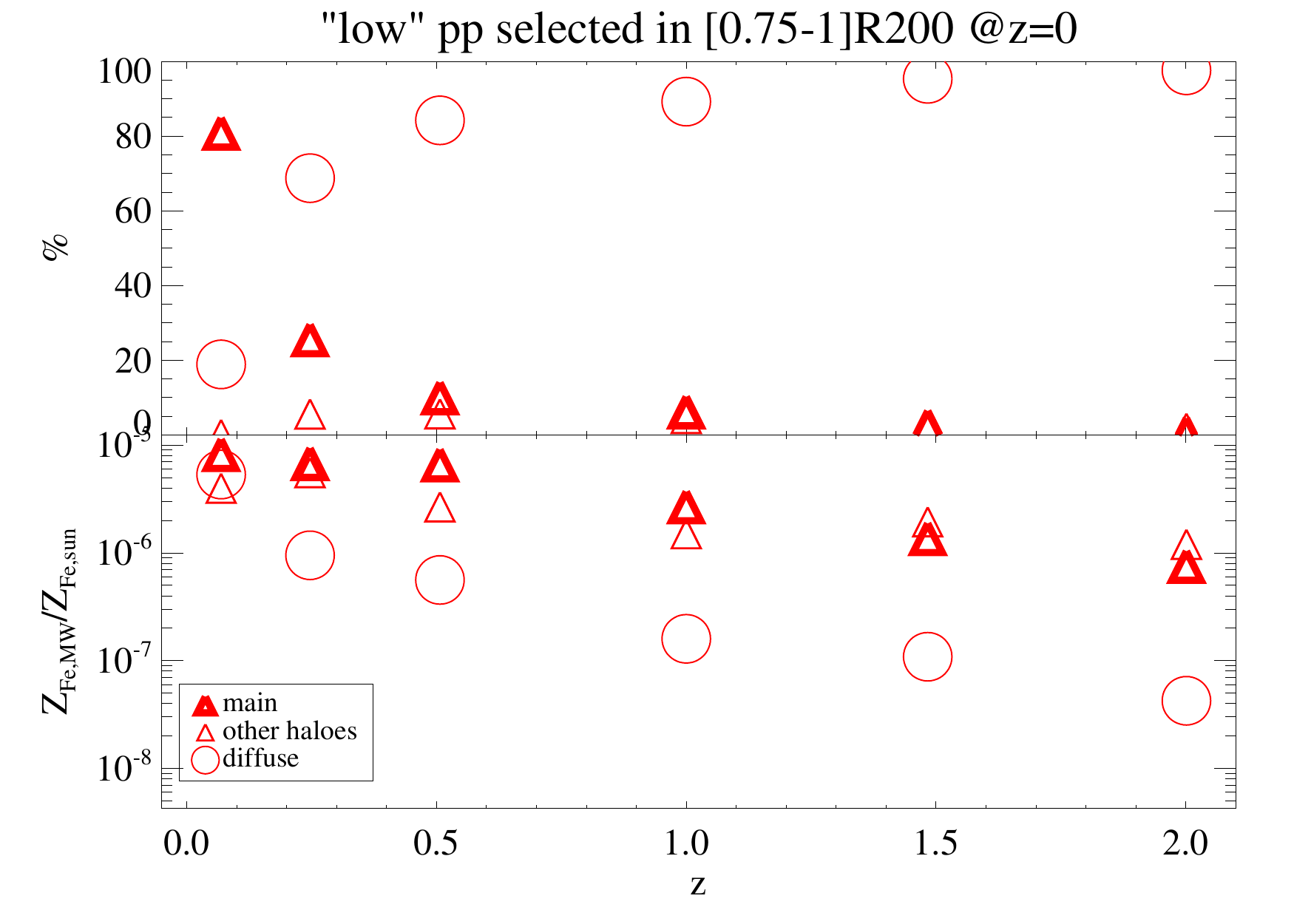}}
  \caption{From top to bottom, same as \protect\figref{fig:pp_inRvir_all} for the tracked
    \hi, \me\ and \lo\ gas subselections, respectively.
    \label{fig:pp_inRvir_Fe}}
\end{figure}

  Similarly to \figref{fig:pp_inRvir_all}, we can distinguish the
  accretion history and chemical evolution depending on the final
  metal content and we show these results for the \hi, \me, and
  \lo\ gas components separately in \figref{fig:pp_inRvir_Fe} (from top
  to bottom, respectively). Also in this case, we explore the spatial
  origin of the tracked particles and quantify whether at any given
  redshift up to $z=2$ they were residing already within the main
  progenitor, within any other surrounding halo more massive than
  $10^{11}\msunh$ or rather outside of them all.

From the upper inset in each panel we can infer that the accretion
history of these three different components has been different.
Namely, the gas that is very poorly enriched at $z=0$ (\lo\ subsample,
in the bottom panel) has been accreted more recently to the main halo,
being for the major part in the diffuse component (more than $\sim
80\%$ down to redshift $z=0.25$).  This is the pristine gas that fills
the space between the haloes and which has been mostly far from active
star formation regions where metal pollution occurs. In fact, its
typical Fe abundance is one order of magnitude smaller
when it is still in the diffuse component than when it is
already included within one halo, either main or neighbouring (see
\figref{fig:pp_inRvir_Fe-lo}, lower inset).
In summary, it has a very poor metal content at $z=0$, and
has been accreted only recently.

As for the gas with Fe abundance close to the median of the
distribution at $z=0$, we note from \figref{fig:pp_inRvir_Fe-med} that
its accretion onto the main halo has been smoother over time.  A small
percentage was residing in the main progenitor already at redshift
$z>1$, and a similarly small percentage was also distributed among the
surrounding haloes.  This increases up to $\sim20\%$ by $z\sim1$, as
this gas gets accreted, and then continues to increase for the main
progenitor only, as the surrounding haloes themselves merge onto the
main cluster at lower redshifts, i.e.\ $z\lesssim 0.5$. This
component, eventually populating the broad peak in the $z=0$
distribution around $\sim 10^{-3}$--$10^{-2}\,Z_{\rm Fe,\odot}$ (see
\figref{fig:distrib}), continues to increase its metallicity from very
low values at $z=2$ ($\sim$ few $10^{-5}\,Z_{\rm Fe,\odot}$ if diffuse, or
$\sim$ few $10^{-4}\,Z_{\rm Fe,\odot}$ if already included within some
halo). This modest level of metal enrichment can mostly be ascribed
to the pollution from long-living stars or residual star formation
episodes during the cosmic time between $z=2$ and $z=0$, rather than
from enrichment within the core of these haloes during the intense
peak of star formation activity at $z\sim 2$--$3$.

A different, interesting picture is outlined by the inspection of the
highly-enriched gas component (\hi\ subselection), shown in
\figref{fig:pp_inRvir_Fe-hi}.  This gas component appears to be
accreted onto the main cluster
 more smoothly in time than the poorly enriched gas, whose
  majority remains outside the main cluster till very low redshifts.
Nevertheless, still $\sim 30\%$ of this
gas is not yet enclosed within any massive halo at $z=0.07$. At higher
redshift, the difference in the fraction of gas residing in the outer
space and in some halo is not as striking as in the other two
(\me\ and \hi) cases.
At $z=2$, if enclosed within some halo, it was mostly within
intermediate-mass surrounding haloes ($\sim 60\%$ of it) rather than within
the main one ($\sim \mbox{few}\%$).
Even more interestingly, we observe that this gas was already
significantly enriched with iron at high redshift, with typical MW Fe
abundances of $\sim 0.4$--$0.5 Z_{\rm Fe,\odot}$ at $z=2$, if enclosed
within some halo.
At $z=2$ almost 40\% of it was not enclosed within any halo with
mass $\mvir > 10^{11}\msunh$, and yet its MW Fe abundance was
already $Z_{\rm MW,Fe}\sim 0.35 Z_{\rm Fe,\odot}$.
This unbound and enriched gas component was most likely
ejected in an earlier phase of AGN activity and then expelled by its galaxy.
While the percentage of the gas outside of any of the haloes
considered does not diminish significantly, its typical $Z_{\rm
MW,Fe}$ increases from $0.35 Z_{\rm Fe,\odot}$ up to $>0.7Z_{\rm
Fe,\odot}$, driven by late enrichment, most likely due to SNIa
rather than fresh star formation generating SNII episodes.

If we compare the results displayed in \figref{fig:pp_inRvir_Fe} with
those in \figref{fig:pp_inRvir_all}, we infer that the Fe-abundance
distribution of the outer gas selected at $z=0$ changes going back in
time.  On the one hand, we find that the most-enriched
particles (\hi) were already Fe-rich at high $z$, both inside and
outside the main and surrounding haloes (see \figref{fig:pp_inRvir_Fe-hi}). On the other hand, the majority of all the
particles had low $Z_{\rm Fe}$ and were residing outside the main halo.
Indeed, of all the particles tracked, we see from
\figref{fig:pp_inRvir_all} that only $\sim 20\%$ of them was already
within the main halo or in a surrounding one at $z=2$, and they had in
both cases MW $Z_{\rm Fe}$ already similar to that at $z=0$. Contrarily,
the remaining $\sim 80\%$ was still to be accreted and had a MW
average iron abundance a factor of $\sim3$--$4$ lower.
Also among the gas in the diffuse component, however, there
was already a small fraction of gas that was highly enriched, as
visible from the backward tracking of the \hi\ subsample, but the
MW-$Z_{\rm Fe}$ is overall diluted by the low-metallicity gas.

\subsection{Iron abundance evolution of the gas in the $z=0$ outskirts}
\label{sec:distrib-evol}

\begin{figure}
  \centering
  \includegraphics[width=0.45\textwidth,trim=10 0 20 25,clip]{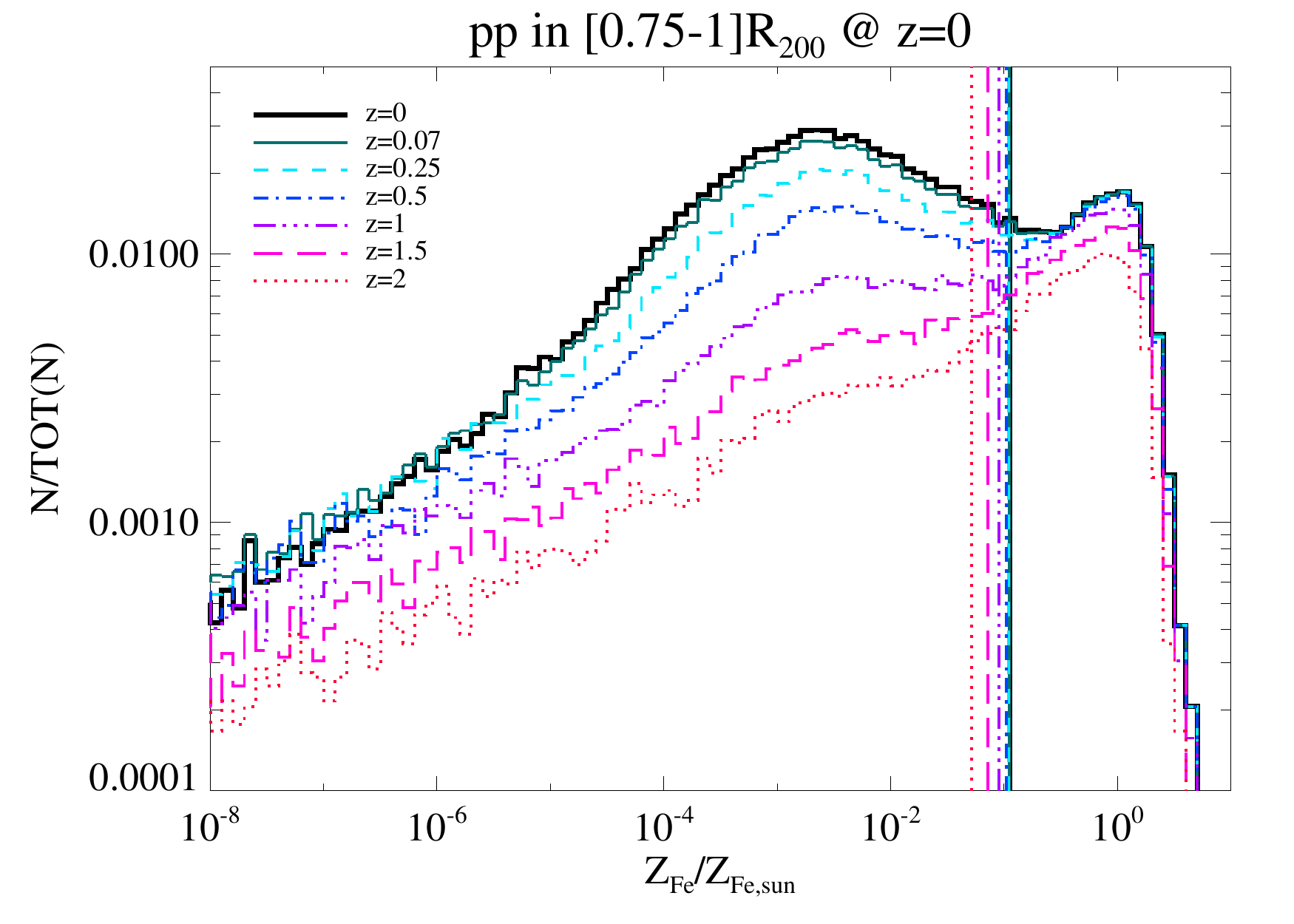}
  \caption{Distribution of the Fe abundance (w.r.t.\  solar values
    by~ANGR89) for the hot-phase gas particles selected to reside within
    $[0.75$--$1]\rtwo$ at $z=0$ and tracked back in time, at various
    redshifts up to $z=2$. Vertical lines correspond to the MW value
    of each distribution.
    \label{fig:Zfe_distrib_evol}}
\end{figure}

The evolution of the $Z_{\rm Fe}$ distribution for all the gas selected in
the cluster outer shell at $z=0$ and tracked back to $z=2$ is shown in
\figref{fig:Zfe_distrib_evol}. From the figure we note that the
high-metallicity peak around solar values is indeed already present at
high redshift, albeit lower by a factor of $\sim2$ in normalization,
whereas the strongest difference concerns the low-metallicity tail of
the distribution.  In particular, the broad peak around
$Z_{\rm Fe}/Z_{\rm Fe,\odot}\sim 10^{-3}$ vanishes towards higher redshifts,
where a greater percentage of the tracked gas has in fact very low
iron abundance or is not enriched at all. We verified that the
percentage of zero-metallicity gas ($Z_{\rm Fe}=0$) increases going back
in time, from $\sim10\%$ at $z=0$ to $\sim80\%$ at $z=2$.
This broader peak corresponds to the average pollution of
  gas particles with iron, and essentially results from the
  combination of the yields and of the distribution of metals from the
  stars to neighbour gas particles according to the SPH kernel
  (i.e. depending on their distance from the stellar source).

\begin{figure}
  \includegraphics[width=0.45\textwidth,trim=10 0 20 25,clip]{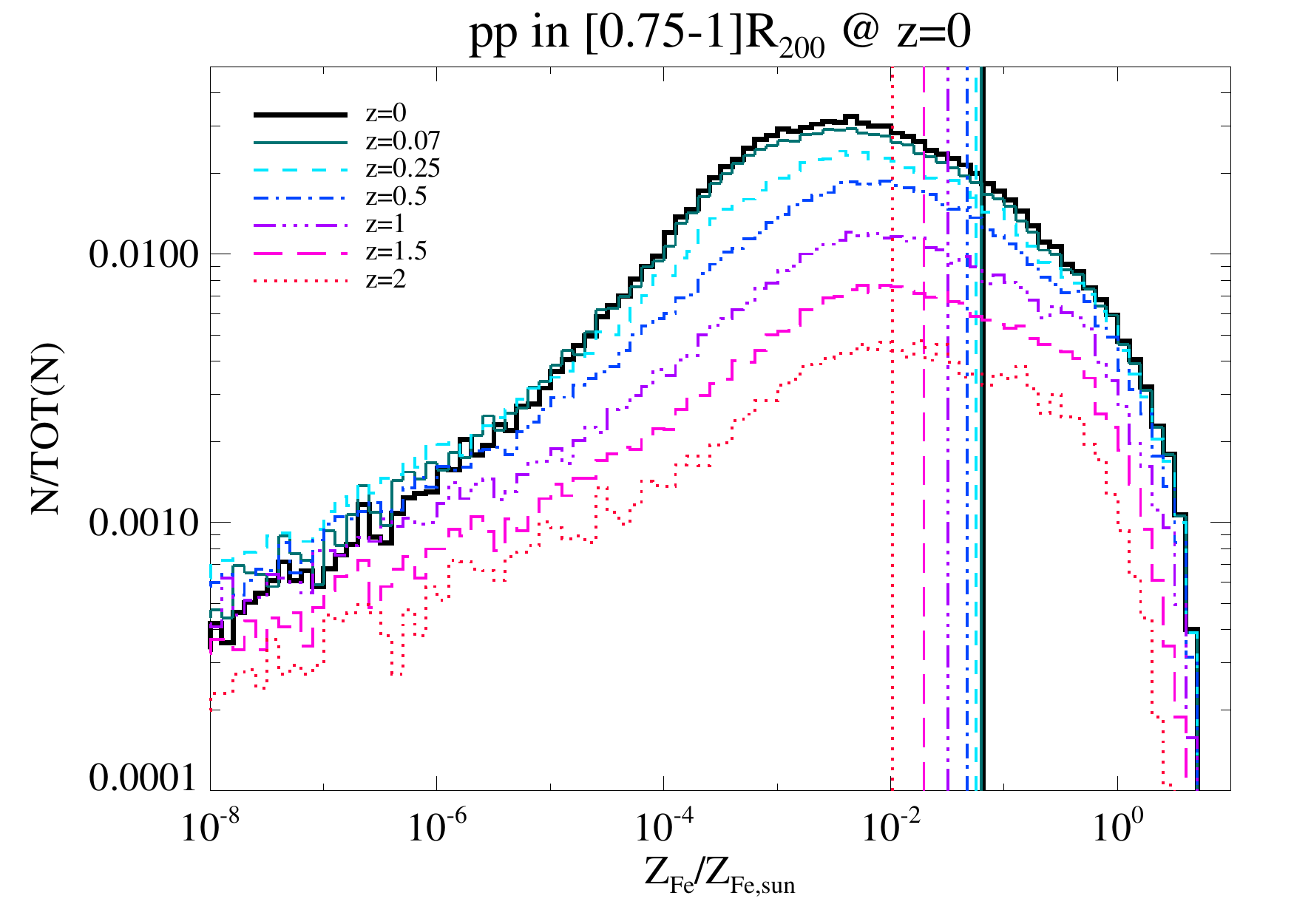}
\caption{Similar to \protect\figref{fig:Zfe_distrib_evol} but for the
  CSF simulation of D2.\label{fig:Zfe_distrib_evol_CSF}}
\end{figure}

Concerning the peak around solar values of $Z_{\rm Fe}$,
the presence of a highly-enriched gas component in the cluster
outskirts at $z=0$, that has clearly been enriched at $z>2$, is a key
imprint of early AGN feedback episodes.
We remind the reader that in our models we do not include a kinetic
feedback so this phenomenon is entirely due to the buoyancy of gas with freshly boosted energy,
that is able to displace metal-rich gas out to large
distances from the star formation regions, since high-redshift
small-mass haloes have shallower potential wells than present-day
massive clusters~\cite[see also][]{Biffi_2017}.
In fact, a similar analysis on the CSF simulation of
the same cluster showed a different distribution for the gas selected
in the same outer shell at $z=0$, displayed in \figref{fig:Zfe_distrib_evol_CSF}.
Namely, no secondary peak of the iron abundance distribution is present around
solar values of $Z_{\rm Fe}$, neither at $z=0$ nor at higher
redshifts. This indicates that the most-enriched gas is typically
confined in the very vicinity of star formation regions and is
preferentially converted into newly formed stars, where the majority
of the metal content is therefore locked. The lack of a powerful
feedback mechanism at high redshift essentially prevents the gas to be
moved away before it is again processed into stars.  In fact, we
verified that a similar tracking of the most enriched gas at $z=0$, in
the CSF case, shows lower iron abundances, by a factor of $2$--$4$ if
enclosed within some halo, throughout the redshift range
$z=0$--$2$. If we track the \hi\ gas outside any halo more massive
than $10^{11}\msunh$, then the typical $Z_{\rm MW,Fe}$ in the \csf\ run is
up to one order of magnitude lower at high redshift.
In fact, as previously noted for \figref{fig:maps0_CSF}, this
behaviour can also be appreciated from the maps of MW iron abundance
in \figref{fig:maps_CSF}, for the $z=0$ \hi\ gas component. The
same spatial scale and color code used for the \agn\ case emphasize
how the most iron-rich gas distribution is clumpier and
characterised by a
lower values of $Z_{\rm MW, Fe}$.  From both \figref{fig:maps0_CSF} and
\figref{fig:maps_CSF}, we see that also in the \csf\ case the metal
distribution traces the large scale structure around the main cluster
progenitor and the accretion of the neighbouring haloes onto it, but
the level of iron abundance is typically lower, with the peaks
of high abundance particularly localized. This difference is remarkable
especially at high redshift, where on the contrary the iron-rich gas
in the \agn\ simulation, as well as the tracked \hi\ component, was
already widely distributed across the whole region connecting the main
cluster and the surrounding haloes.



\section{Tracing the contribution from different enrichment sources}\label{sec:track}

In this section we will discuss the origin of the metal content
of the gas residing in the cluster outer shell at $z=0$
in terms of the {\it contribution to the enrichment from SNIa and SNII}.

In fact, SNII release mainly light metals (such as, O,
  Ne, Mg or Si), whereas heavier elements (like Fe and Ni) are
  predominantly produced by Type Ia supernovae. Iron is in reality
  produced by both SNII and SNIa, but with different time-scales and
  yields, being the contribution of SNIa the most significant
  especially at later times.  Lighter elements such as C or N are produced
  mainly by low- and intermediate-mass AGB stars.

\begin{figure}
  \centering
  \includegraphics[width=0.45\textwidth,trim=15 0 15 22,clip]{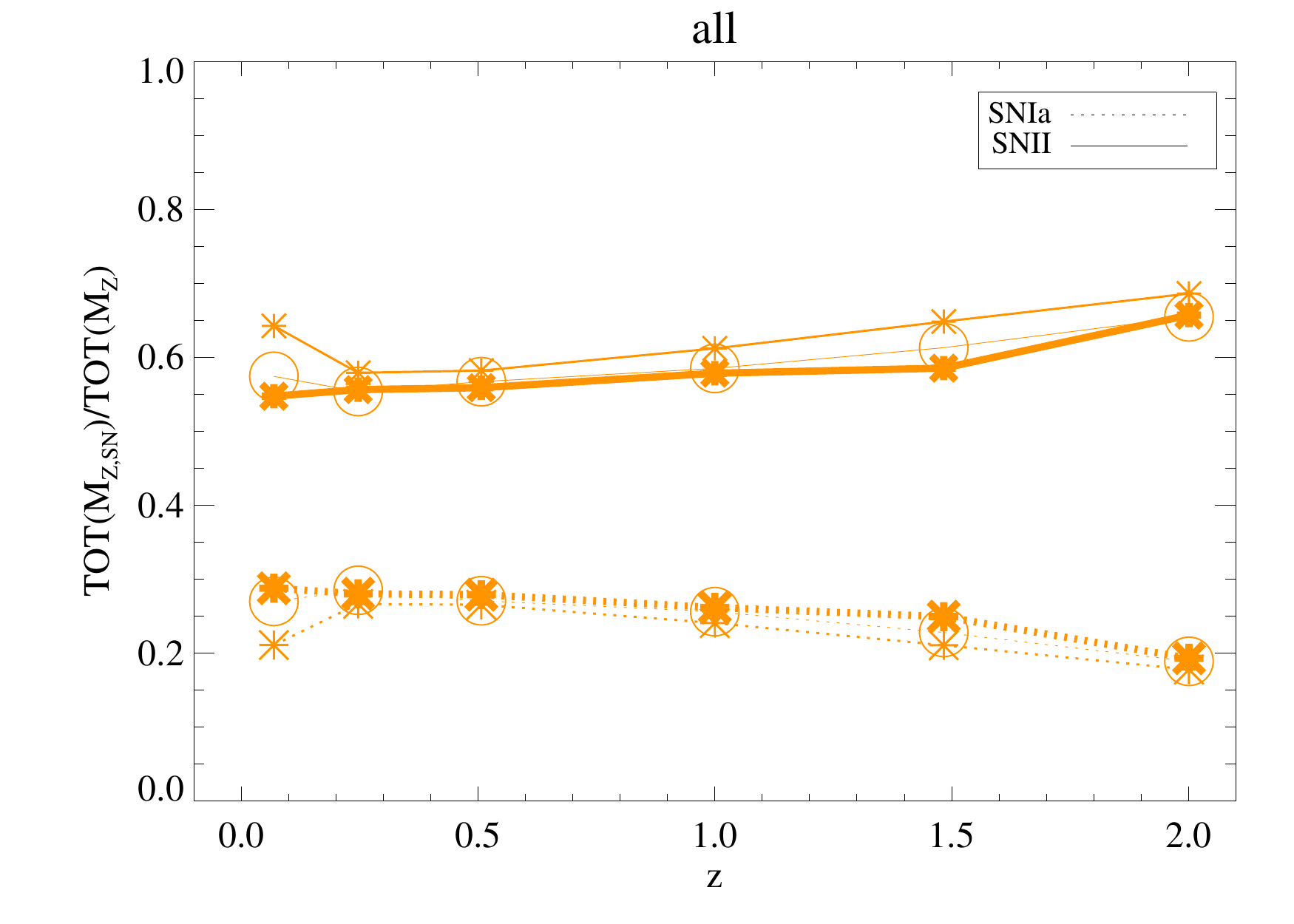}
  \caption{Fraction of the mass of metals produced by SNIa (dotted
    lines) or SNII (solid lines) with respect to the total mass of
    metals, for the gas selected at $z=0$ in the spherical shell
    $[0.75$--$1]\,\rtwo$ and tracked back in time up to $z=2$.
    Different symbols refer to the location of the tracked gas at
    redshift $z$, namely within the main progenitor (thick
    asterisks), within the virial radius of any other surrounding halo
    with $\mvir > 10^{11}\msunh$ (thin asterisks), or outside of them all (empty
    circles).
    \label{fig:trck-Ia-II-all}}
\end{figure}

Throughout the cosmic history, different stellar sources contributed
to producing metals and polluting the surrounding gas. The specific
contribution from SNIa and SNII -- and consequently of the AGB stars
-- can be evaluated directly in our simulation.
In \figref{fig:trck-Ia-II-all} we show the metal mass fraction due to
SNIa (solid lines) and SNII (dotted lines) for all the metal-rich gas
tracked at various redshifts between $z=0$ and $z=2$.
As expected, the contribution from SNII is dominating the
metal mass fraction at all redshifts.  This means that the majority of
metals, in terms of total mass, is produced by SNII,
which is in fact consistent with oxygen being the most abundant
element in the Universe.  On the other hand, the mass of metals
produced by SNIa increases with decreasing redshift.  In fact, the
evolution with redshift of these contributions is opposite and in
agreement with the typical life time scales associated to the two
stellar sources: overall, the SNIa-fraction increases from $\sim 20$ per cent
at $z=2$ to $\sim 30$ per cent at $z=0.07$; differently, the SNII-fraction
decreases from $\sim 65$ per cent to $\sim 55$ per cent in the same temporal
range. From \figref{fig:trck-Ia-II-all}, one can easily infer that the
remaining metal mass is due to AGB stars, whose contribution roughly
oscillates around $\sim 15$ per cent.
The different symbols used (as in \figref{fig:pp_inRvir_all}) refer to
the location of the tracked gas at the given redshift $z$, so that we
can differentiate among that already accreted onto the main cluster,
the fraction still enclosed within some other intermediate-mass
neighbouring halo and the fraction that is still outside of them all.
Interestingly, we find that the trend does not depend significantly on
the environment in which the gas is located, suggesting that this
result is mainly driven by the time-delay and yields of the
enrichment channels followed by our simulations.  Indeed, variations among
the three sets of values are on average within $10\%$.

\begin{figure}
  \centering
  \subfigure[\hi\ gas.\label{fig:trck-Ia-II-hi}]{\includegraphics[width=0.45\textwidth,trim=15 0 15 22,clip]{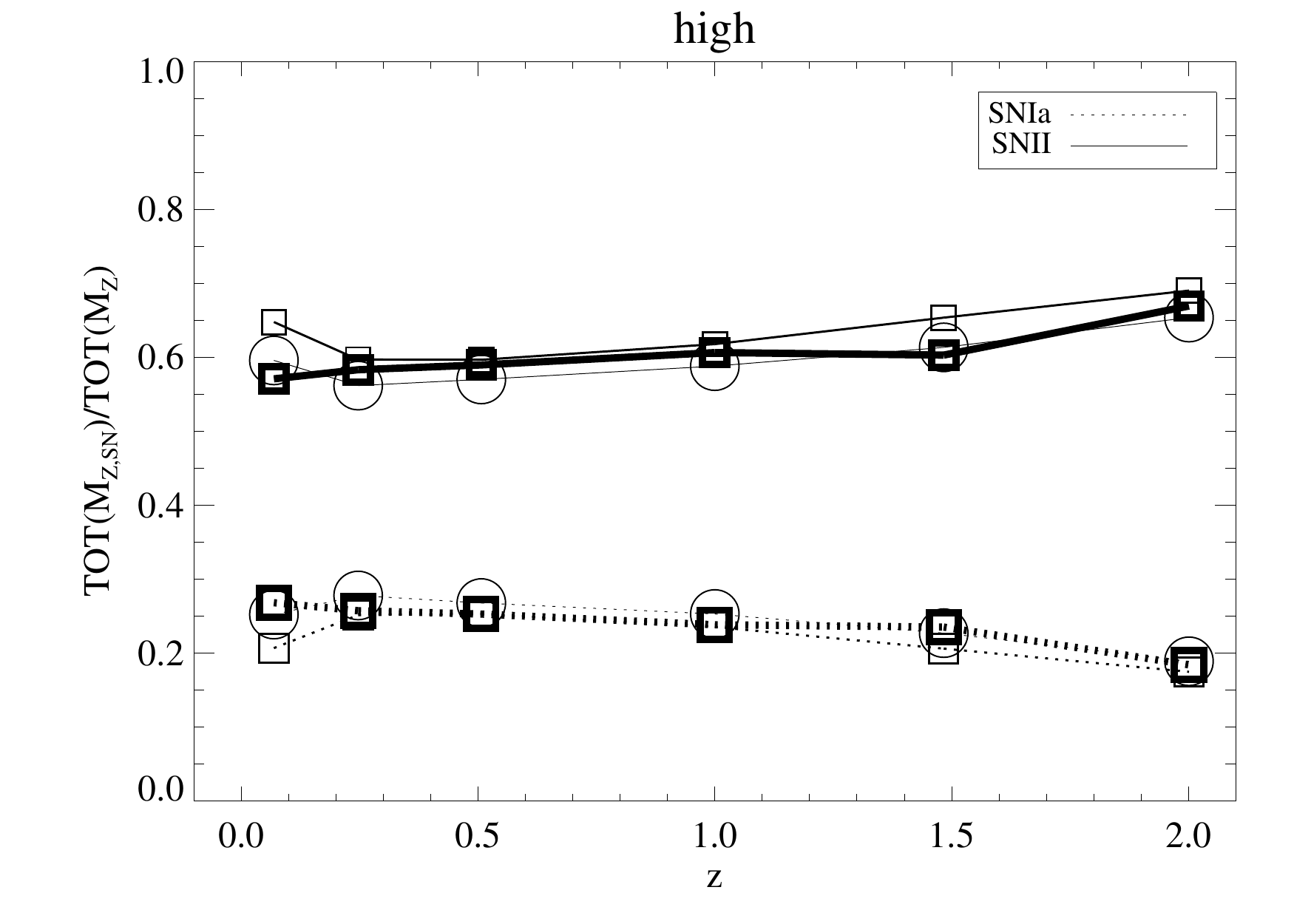}}
  \subfigure[\me\ gas.\label{fig:trck-Ia-II-med}]{\includegraphics[width=0.45\textwidth,trim=15 0 15 22,clip]{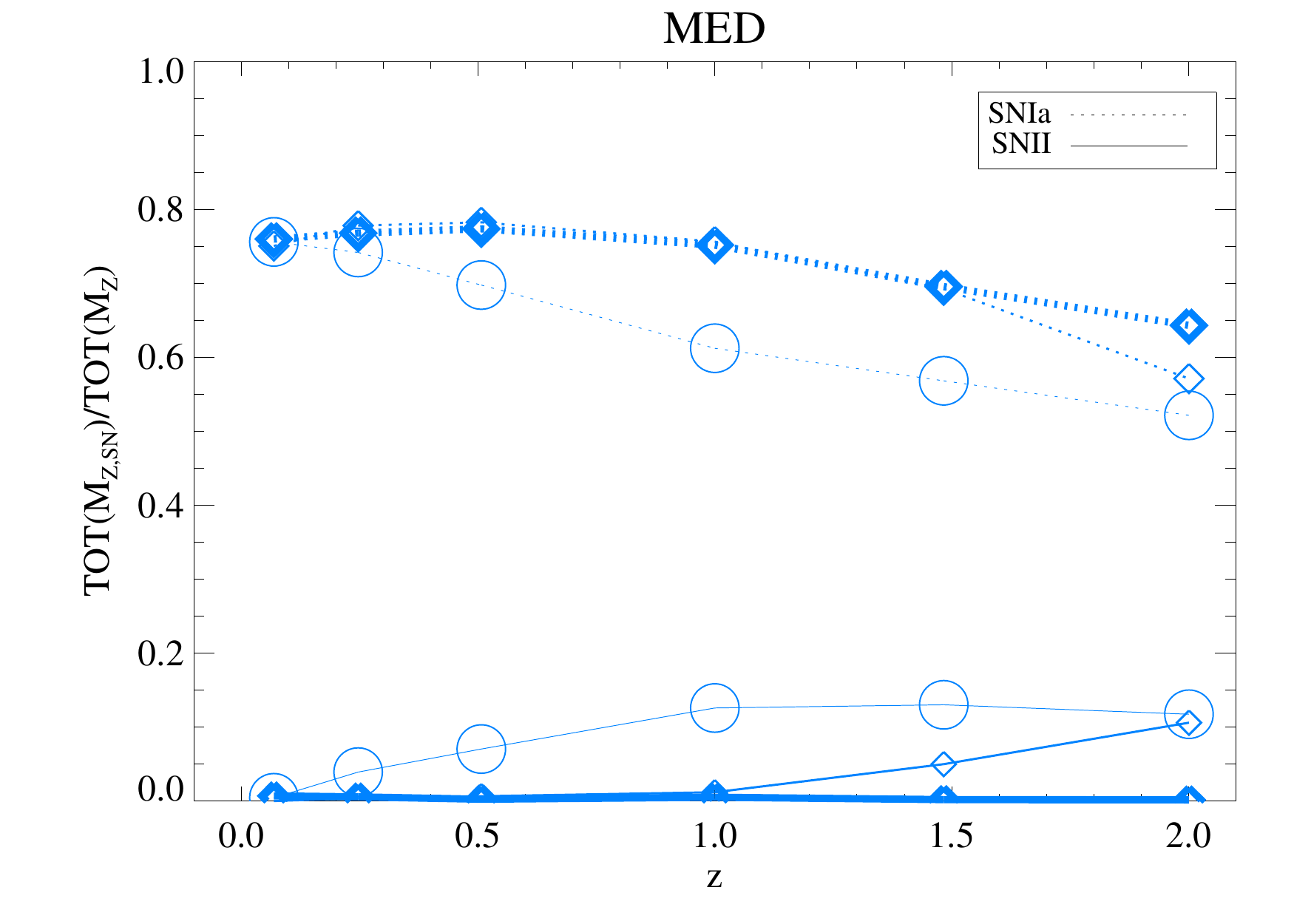}}
  \subfigure[\lo\ gas.\label{fig:trck-Ia-II-lo}]{\includegraphics[width=0.45\textwidth,trim=15 0 15 22,clip]{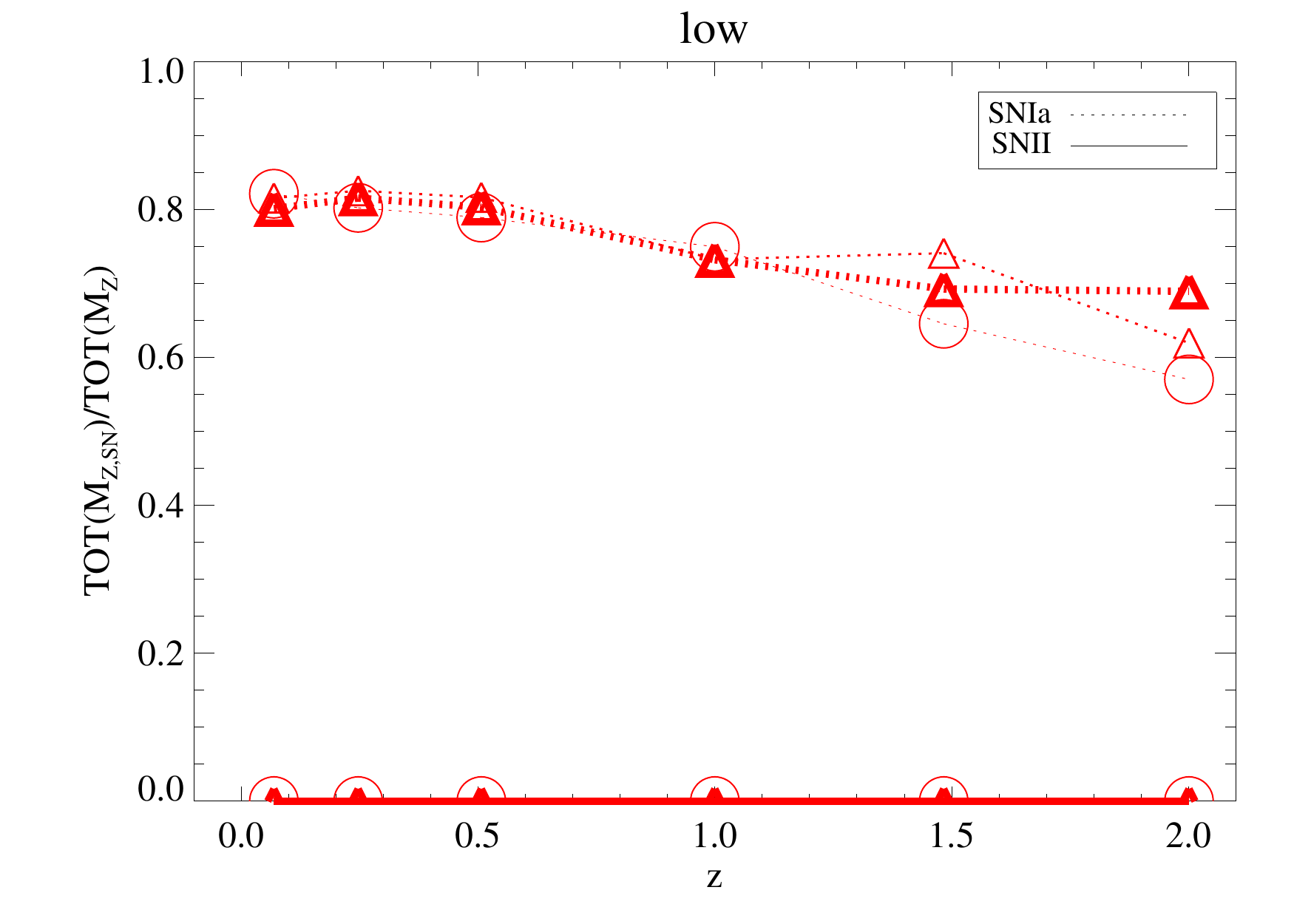}}
  \caption{From top to bottom, same as
    \protect\figref{fig:trck-Ia-II-all} for the tracked \hi, \me\ and
    \lo\ gas subselections, respectively. (Symbols refer to different
    environments as in Figs.~\protect\ref{fig:pp_inRvir_all},
    \protect\ref{fig:pp_inRvir_Fe} and
    \protect\ref{fig:trck-Ia-II-all}).
    \label{fig:trck-Ia-II}}
\end{figure}

In \figref{fig:trck-Ia-II} we show the contribution from these two
main enrichment channels for the subselections of the gas based on its
$z=0$ iron abundance (\hi, \lo\ and \me\ components, from top to
bottom, respectively).%
The \hi\ gas component shows very similar results to those discussed
for the whole gas in \figref{fig:trck-Ia-II-all}. This suggests that
the high-$Z_{\rm Fe}$ gas -- which has been found to have relatively high
$Z_{\rm Fe}$ also at higher redshifts -- essentially dominates the mass
budget, namely contains the majority of metals in mass. This gas must
have been therefore significantly enriched by both SNIa and SNII,
since very early times.

Differently, the gas with median or very low Fe abundance at $z=0$,
which has typically even lower abundances at higher redshifts, is
characterised by a metal mass content, albeit modest, which is
essentially dominated by the contribution from SNIa.
In both cases, especially at late times, the SNIa-channel is
contributing up to 80\% of the total metal mass because of the longer
time-scales of their enrichment.
The contribution from SNII is basically negligible, except for a 20\%
contribution to the enrichment of the \me\ gas at $z\gtrsim1$,
probably within haloes smaller than those considered here (i.e.\ with
$\mvir < 10^{11}\msunh$).
Instead, the remaining fraction of metal mass (comprised between
$20\%$ at late times and $\sim40\%$ at early times) is due to AGB
sources.

SNIa, and in a minor proportion AGB stars, essentially contribute to
build up the broad peak centered around $10^{-3}\,Z_{\rm Fe,\odot}$, in
the distribution shown in \figref{fig:Zfe_distrib_evol}.
This all indicates that the poor enrichment of this gas happened far
from SF regions, where the pollution due to short-living SNII is not
significant. To a closer inspection, we found in fact that the
distribution of the metal-rich particles that we track shows a
significant component of quasi-pristine gas that contains only a little
amount of metals almost entirely coming from SNIa sources. This gas
has very low $Z_{\rm Fe}$ and has been probably enriched by late SNIa
episodes far away from SF regions, given that long-living stars have
more time to move away before dying as SNIa and polluting
the surrounding medium.

Both \figref{fig:trck-Ia-II-all} and \figref{fig:trck-Ia-II} remark that
the properties of the tracked gas with respect to the enrichment sources are
essentially the same independently of whether it is residing within
a halo (either the main or a neighbour one) or not, from $z=2$ down to $z\sim 0$.
  The results obtained with our current modelling of chemical evolution
  and AGN thermal feedback, further support
the idea that the bulk of
the enrichment is not happening ``in situ'' within the formed
clusters, but is rather the result of a more complex and continuous
process of accretion of both pristine gas and highly metal-rich gas,
that was previously enriched within neighbouring
substructures that eventually merge onto the main halo.


\section{Summary and conclusion}\label{sec:conclusions}

In this paper we focused on the origin of ICM chemical enrichment
in the present-day outskirts of simulated clusters
from a set of cosmological, hydrodynamical zoomed-in simulations of
galaxy clusters performed with an improved version of the GADGET-3
code.
The simulations include the treatment of a variety of
  physical processes describing the physics of the baryonic component,
  among which thermal feedback from AGNs.
Results on the thermo- and chemo-dynamical properties of
clusters in these simulations have been shown to agree with a variety
of observational evidences in a series of recent
papers~\cite[][]{rasia2015,villaescusa2016,truong2018,Biffi_2016,
  Planelles_2017,Biffi_2017}.
Here, we traced back in time the hot gas particles that reside in the
outskirts of the cluster at $z=0$, defined as the region between
$0.75\rtwo$ and $\rtwo$, and investigated the evolution of their
spatial distribution and enrichment level and source, up to $z=2$.  We
remark that the properties
that we discuss are always referred to the gas residing in the
present-day outskirts and tracked back in time, only.

With the present analysis we aim at the detailed origin of the
remarkably uniform metal enrichment typical of present-day cluster
outskirts, which is found both in
observations~\cite[][]{werner2013,simionescu2015,urban2017}
and in simulations~\cite[][]{Biffi_2017,vogelsberger2017}.
From simulations, we find that the flat metallicity and abundance
profiles at large cluster-centric distances turn into steeper and
lower-normalization gradients when exclusively stellar feedback is
included in the simulations. This is a footprint of the crucial role
played by high-redshift AGN feedback in distributing the metal-rich
gas far away from star formation regions, out and much beyond the
virial radius of surrounding haloes in the proto-cluster environment.

While we presented results for one single study-case cluster, we find
that very similar trends are confirmed when different clusters, in
terms of mass and cool-coreness, are investigated. This is discussed
in Appendix~\ref{app:others}.  In particular, we find that the key
feature of the present-day iron abundance distribution of the hot gas
in the outer cluster shell, namely the secondary peak around solar
values, is always present in the four clusters analysed, and already
from redshift $z=2$.

We summarize hereafter our main findings.

\begin{itemize}

\item A component of the gas selected in the present-day outer shell
  of the (\agn) cluster is highly enriched with iron, with $Z_{\rm Fe}$
  peaked around solar values. This component dominates the
  mass-weighted average Fe abundance. We
  find that this component is also already present at $z=2$ and
  already with significant enrichment level, while the fraction of
  median-$Z_{\rm Fe}$ gas increases, due to the later enrichment.

\item Most of the gas in the present-day outskirts has been accreted
  onto the main cluster only recently. At $z=0.5$, about $\sim 70\%$
  of the gas found in these regions was not yet accreted onto the main
  halo. At redshift $z\ge 1$ more than $60\%$ of this gas
  is in the diffuse component or in much smaller haloes.

\item The mass-weighted iron abundance typical of this tracked gas is
  comprised between $0.1$ and $0.2$ solar, and does not evolve much as
  long as it is residing in any halo, either main or not, at higher
  redshifts. The gas in the diffuse component increases its MW
  $Z_{\rm Fe}$ only by a factor of a few (from $0.02\,Z_{\rm Fe,\odot}$ to
  $0.15\,Z_{\rm Fe,\odot}$), mainly due to the fact that the diffuse
  component at high redshifts is diluted by a significant fraction of
  pristine gas.

\item From the backward tracking of the gas particles, our results
  indicate that the highly enriched gas residing in the outskirts at
  $z=0$ was already significantly enriched at higher redshifts up to
  $z=2$ (when, $Z_{\rm Fe}\sim0.4$--$0.5\,Z_{\rm Fe,\odot}$).  This gas has
  been likely enriched at even higher redshifts, during or before the
  peak of star formation, and was later accreted onto the main
  cluster, together with more pristine gas, both from the diffuse component
  and through the merging with surrounding haloes during the cluster
  mass assembly.

\item The chemical enrichment of the tracked gas is overall dominated,
  in mass, by the metal pollution due to SNII ($\sim 55$--$65\%$, from
  redshift $z=0$ to $z=2$). This is essentially driven by the
  enrichment characteristics of the most metal-rich gas (for which the
  SNII metal mass fraction is $\sim 60$--$70\%$, in the range
  $0<z<2$). There is however a gas component, the one with median or very
  low iron abundance, which is mainly enriched by SNIa ($\sim
  60$--$80\%$, from redshift $z=0$ to $z=2$) -- and partially AGB
  ($\sim 20\%$ at $z=0$ to $30$--$40\%$ at $z=2$) -- products,
  indicating that the enrichment, albeit modest, has happened at
  later times and mostly far from SF regions.

\item From the study of the enrichment sources, we note that this
  shows no significant dependency on the environment where the tracked
  gas resides at a given redshift. Relative proportions of metal mass
  fractions for SNIa and SNII are the same for both the gas within
  some halo (either main or neighbour) or in the diffuse component.

\item When we consider the simulation of the same cluster that does
  not include AGN feedback but only stellar feedback (\csf), we
  observe a sparser distribution of the Fe-rich gas, when the selected
  particles at $z=0$ in the outskirts are traced back in time. Also,
  their typical metal enrichment level is lower and, in fact, their
  distribution of iron abundances shows no component peaked at solar
  or higher values, either at $z=0$ where they are selected or at $0
  < z \le 2$.

\end{itemize}

The analysis presented in this paper confirms that the relatively high
and uniform metallicity of the ICM in the outskirts of nearby clusters,
traced by its iron abundance, is the result of the accretion with time
of both pristine and already highly enriched gas, from both the
diffuse component and from surrounding haloes in the proto-cluster
environment that later merge onto the progenitor of the main cluster
during its formation and mass assembly. After the
high-redshift pollution of the gas with metals, by both SNII and SNIa,
mixing and dynamical processes have played a major role in
distributing them and in shaping the present-day profiles at large
cluster-centric distances.
Interestingly, we observe in simulations that at $z=2$ part of the
highly enriched gas is diffuse and resides outside of any
intermediate-mass halo in the protocluster region. This gas has been
expelled out to large distances from the pollution sites at higher
redshifts by powerful AGN episodes in small-mass haloes. This can be
inferred by the comparison to simulations including only stellar
feedback, where in fact the level of ICM enrichment is lower and much
more localized, confined to the SF sites~\cite[see also][]{Biffi_2017}.

While the present-day imprints of this pre-enrichment mechanism has
already found confirmation in observational evidences, the ideal next
step would be to push X-ray observations of the ICM enrichment to the
region close or immediately beyond the virial radius or to the diffuse
medium between pre-merging systems, and to clusters at higher
redshifts.  In fact, this will help constraining the widespread
distribution of metal-enriched gas in the proto-cluster
environment~\cite[see also studies on galaxies in dense
  proto-cluster regions at high redshifts, e.g. ][]{overzier2016}, and
the later accretion of these metals to the cluster, rather than the
in-situ production at lower redshifts.
To this goal, next generation of X-ray satellites with large effective
areas and high-resolution spectroscopy, such as
\athena~\footnote{http://www.the-athena-x-ray-observatory.eu/}~\cite[][]{athena,Barret_2016},
will enable far more detailed observations of the cluster outskirts and
reliable measurements of the ICM metallicities.

From the point of view of numerical simulations, we remark here that a
number of physical processes -- such as dust production and disruption
or metal diffusion -- are not yet included in this set of simulations
and will need to be accounted for in order to pursue further detailed
studies on the chemical evolution of galaxy clusters.
An additional limitation of the current simulation set is the
  resolution, which prevents us from resolving the details of the
  galaxy component in the clusters, especially at high redshift.
  Although the ICM pre-enrichment picture presented here is consistent
  with results from independent studies on higher-resolution simulation
  sets~\cite[see recent results by][]{vogelsberger2017},
  as a future improvement, we plan to perform and investigate
  higher-resolution simulations in order to consistently explore the
  details of the galaxy population within clusters, which interact
  with the ambient ICM since high redshift and contribute to both
  energy and chemical feedback~\cite[see e.g.][]{renzini2014}.
  This will allow us to study the details of the star formation
  history and chemical patterns of the stellar component within
  galaxies, adding information to the picture discussed in the present
  analysis.

Combining high resolution and improvements on the physical
  description of the simulations
will be crucial especially for detailed comparisons with observational
findings, also of the high-redshift Universe, expected from upcoming
missions like \athena.


\section*{Acknowledgements}
The authors would like to thank V.~Springel for allowing us to access
the developer version of the {\small GADGET} code and L.~Steinborn for
providing access to the improved AGN feedback model.  The authors are
thankful to G.L.~Granato, C.~Ragone-Figueroa, L.~Tornatore and
K.~Dolag for useful discussions during the early stages of this work,
and to the anonymous referee for constructive and insightful
  comments on the manuscript.
Simulations have been carried out using Flux HCP Cluster at
the University of Michigan, Galileo at CINECA (Italy), with CPU time
assigned through ISCRA proposals and an agreement with the University
of Trieste.  We also acknowledge PRACE for awarding us access to
resource ARIS based in Greece at GRNET, through the DECI-13 PRACE
proposal.  The post-processing has been performed using the PICO HPC
cluster at CINECA through our expression of interest.
We acknowledge financial support from the PRIN 2015W7KAWC founded by
the Italian Ministery for University and Research, the INFN INDARK
grant, and ``Consorzio per la Fisica'' of Trieste.
SP acknowledges support from the ``Juan de la Cierva'' program (ref. IJCI-2015-26656) funded by the {\it Spanish Ministerio de Econom{\'i}a y Competitividad} (MINECO) as well as from the MINECO through the grants AYA2013-48226-C3-2-P and AYA2016-77237-C3-3-P and the Generalitat Valenciana (grant GVACOMP2015-227).
DF acknowledges financial support from the Slovenian Research Agency
(research core funding No. P1-0188).
M.G. is supported by NASA through Einstein Postdoctoral Fellowship
Award Number PF5-160137 issued by the Chandra X-ray Observatory
Center, which is operated by the SAO for and on behalf of NASA under
contract NAS8-03060. Support for this work was also provided by
Chandra grant GO7-18121X.

\bibliographystyle{mnbst}
\bibliography{track.bib}

\appendix

\section{Behaviour of different clusters}
\label{app:others}

We performed a similar analysis to the one shown for D2 in the
previous sections on the representative subsample of clusters, for
which the tracking of the metal origin is available, listed in
Table~\ref{tab:glob-prop}.  Far from having
any statistical purpose, the aim of this test is purely to explore
whether dependencies of our results on the system mass or central
thermal properties exist or not.

Observational findings, as well as numerical investigations~\cite[e.g.][]{yates2017,barnes2017,vogelsberger2017,dolag2017}, have
shown that the chemical properties of the ICM at large cluster-centric
distances is extremely uniform across different clusters, and is
essentially independent on the specific thermal properties of the
cluster core, such as cool-coreness, which cause instead a larger
variation in the core, from cluster to cluster.
Here we further explore whether this homogeneity regards as well the
origins of the present-day outskirts metal content.

\begin{figure}
  \centering
  \includegraphics[width=0.47\textwidth,trim=10 0 20 25,clip]{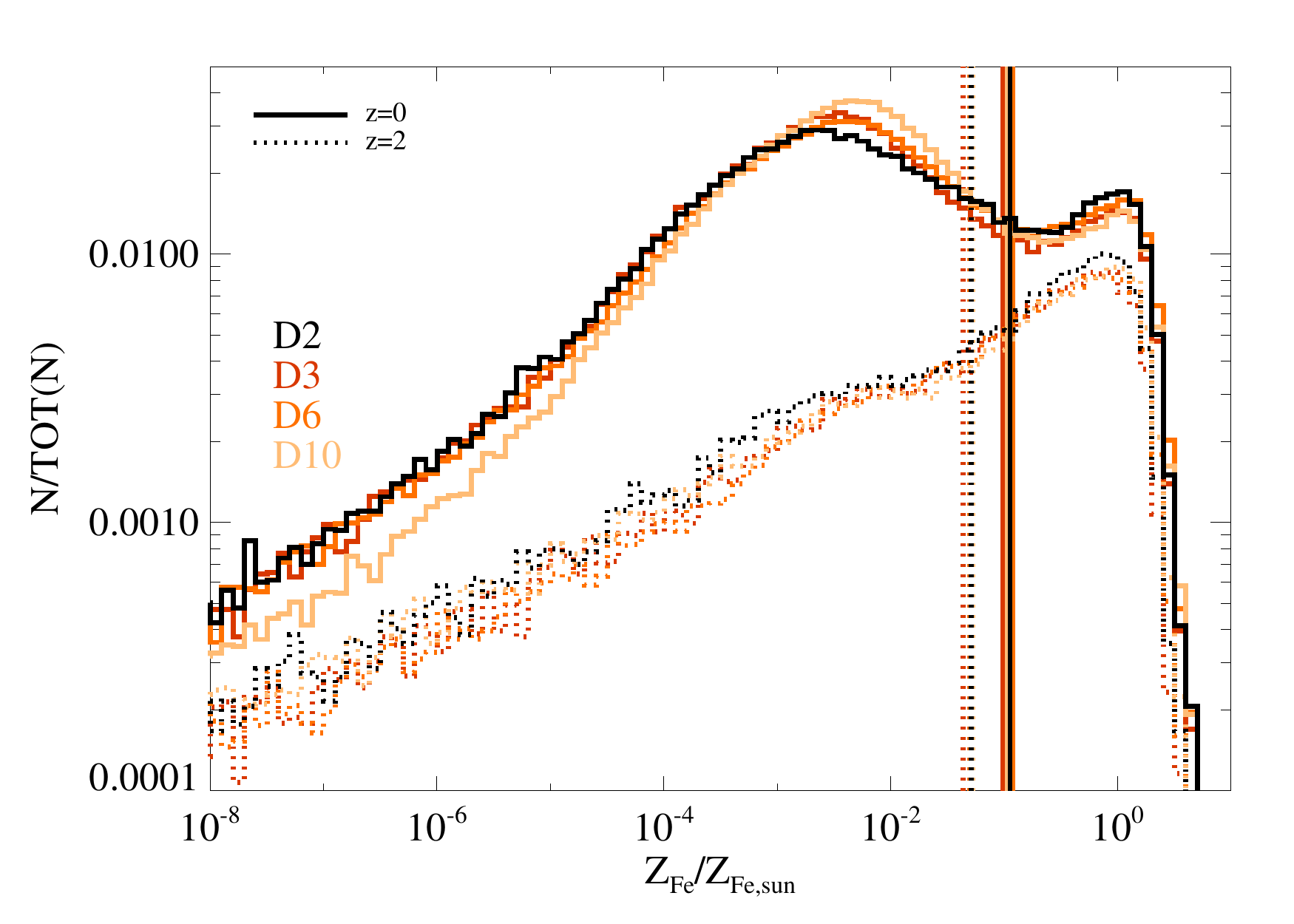}
  \caption{Distribution of the Fe abundance (w.r.t.\ solar values
    by~ANGR89) for the hot-phase gas particles selected to reside within
    $[0.75$--$1]\rtwo$ at $z=0$ and tracked back in time, shown for
    $z=0$ (solid lines) and $z=2$ (dotted lines) for all the four clusters
    as in the legend. Vertical lines correspond to the MW value of
    each distribution.
    \label{fig:Zfe_distrib_evol_allregs}}
\end{figure}

With the same criteria used for D2, we select and track back in time
the gas in the outskirts of the present-day clusters in
Table~\ref{tab:glob-prop} (namely the hot-phase ICM enclosed in the
spherical shell between $0.75\rtwo$ and $\rtwo$) and we find no
remarkable differences between D2 and the other three systems. The
distribution of the iron abundance of the selected gas at $z=0$ and
its evolution at higher redshifts presents in fact the same features
than those presented in Figs.~\ref{fig:distrib}
and~\ref{fig:Zfe_distrib_evol}. More specifically, the characteristic
peak at high iron abundances is found in all the four clusters
(either small or massive, CC or NCC) and is already clearly present at
$z=2$, indicating that that gas was already significantly enriched by
then.
This is visible from \figref{fig:Zfe_distrib_evol_allregs}.%

Furthermore, the MW iron abundance of the tracked gas always moderatly
varies between $0.1$ and $0.2$, if enclosed within some halo in the
region, whereas it increases from $\sim 0.02$--$0.03$ at $z=2$ to
$\sim 0.1$--$0.15$ at $z=0$ for the gas in the diffuse component.
When the gas traced is selected on the base of its present-day iron
abundance, we also find similar results to those discussed for D2.
Also, the contribution to the metal mass budget of the selected gas is
always dominated by SNII, with similar proportions than in D2.  The
iron-poor gas, instead, is also in all these cases primarily polluted
by SNIa at all redshifts, indicating a late and limited enrichment far
from the active star-formation regions.

This further confirms the uniformity of the metal enrichment
in the outskirts of clusters at $z=0$, independently of their mass and
central thermal properties. Also, they support the idea of an early
enrichment to which both SNIa and SNII have contributed, mainly within
small high-redshift haloes from which the metal-rich gas has been
pushed out by early AGN feedback, able to overcome their still shallow
potential wells. The pre-enriched gas was then accreted from both the
diffuse component as well as with the merging of the surrounding
haloes during the formation of the cluster.
Essentially, the backward tracking analysis adds an important piece to
the puzzle: not only the present-day metal distribution of cluster
outskirts, but also the {\it origin of the metals} and their evolution
in the ICM at the periphery of clusters of different mass and thermal
core properties share remarkably similar features.

We note that minor differences among the four clusters analysed,
mainly in the values of MW $Z_{\rm Fe}$ of the tracked gas residing within
surrounding haloes, rather correspond to differences in their
dynamical history and mass assembly. For instance this can be related
to the presence of -- and merging with -- systems, in the
proto-cluster environment, that have comparable mass to the main
progenitor, rather than the accretion of smaller-mass (by a factor of
2 or more) haloes such as in the case of D2.



\end{document}